\documentclass[aps,prd,twocolumn,nofootinbib,showpacs,superscriptaddress]{revtex4-1}

\usepackage{amsfonts}
\usepackage{amsmath}
\usepackage{amssymb}
\usepackage{bm}
\usepackage{dcolumn}
\usepackage{epsfig}
\usepackage{graphicx}
\usepackage{graphics}
\usepackage[latin1]{inputenc}
\usepackage{latexsym}
\usepackage{rotating}
\usepackage{hyperref}
\usepackage[usenames]{color}
\usepackage{xspace} 
\usepackage{mathrsfs}
\usepackage{subfigure}
\usepackage{enumitem}
\usepackage{tabularx}
\usepackage{multirow,hhline}
\usepackage{booktabs}
\usepackage{siunitx}
\usepackage{array}
\newcolumntype{L}[1]{>{\raggedright\let\newline\\\arraybackslash\hspace{0pt}}m{#1}}
\newcolumntype{C}[1]{>{\centering\let\newline\\\arraybackslash\hspace{0pt}}m{#1}}
\newcolumntype{R}[1]{>{\raggedleft\let\newline\\\arraybackslash\hspace{0pt}}m{#1}}
\newlength{\Oldarrayrulewidth}
\newcommand{\Cline}[2]{%
	\noalign{\global\setlength{\Oldarrayrulewidth}{\arrayrulewidth}}%
	\noalign{\global\setlength{\arrayrulewidth}{#1}}\cline{#2}%
	\noalign{\global\setlength{\arrayrulewidth}{\Oldarrayrulewidth}}}

\usepackage{ulem}
\definecolor {darkgreen}{rgb}{0.2,0.7,0.2}
\graphicspath{{../Figures/}}

\newcommand{\be}{\begin{equation}}
\newcommand{\ee}{\end{equation}}
\newcommand\ba{\begin{eqnarray}}
\newcommand\bse{\begin{subequations}}
\newcommand\ea{\end{eqnarray}}
\newcommand\ese{\end{subequations}}

\newcommand{\nn}{\nonumber}

\newcommand{\eq}{\,=\,}

\newcommand{\mat}{{\mbox{\tiny mat}}}

\newcommand{\SC}{{\mbox{\tiny sc}}}

\newcommand{\DEF}{{\mbox{\tiny DEF}}}
\newcommand{\COSH}{{\mbox{\tiny hyp}}}
\newcommand{\J}{{\mbox{\tiny J}}}
\newcommand{\E}{{\mbox{\tiny E}}}

\newcommand{\scm}{{\mbox{\tiny sc,max}}}

\newcommand{\ppN}{{\mbox{\tiny PPN}}}

\newcommand{\bbn}{{\mbox{\tiny BBN}}}
\newcommand{\crit}{{\mbox{\tiny crit}}}
\newcommand{\fail}{{\mbox{\tiny fail}}}
\newcommand{\ffail}{{\mbox{\tiny all fail}}}

\begin{document}
%
\title{
Solar System Constraints on Scalar-Tensor Gravity with Positive Coupling Constant \\upon Cosmological Evolution of the Scalar Field}

\author{David Anderson}
\affiliation{eXtreme Gravity Institute, Department of Physics, Montana State University, Bozeman, MT 59717, USA.}

\author{Nicol\'as Yunes}
\affiliation{eXtreme Gravity Institute, Department of Physics, Montana State University, Bozeman, MT 59717, USA.}

\date{\today}

\begin{abstract} 

Scalar-tensor theories of gravity modify General Relativity by introducing a scalar field that couples non-minimally to the metric tensor, while satisfying the weak-equivalence principle. 
These theories are interesting because they have the potential to simultaneously suppress modifications to Einstein's theory on Solar System scales, while introducing large deviations in the strong field of neutron stars. 
Scalar-tensor theories can be classified through the choice of conformal factor, a scalar that regulates the coupling between matter and the metric in the Einstein frame. 
The class defined by a Gaussian conformal factor with negative exponent has been studied the most because it leads to spontaneous scalarization (i.e.~the sudden activation of the scalar field in neutron stars), which consequently leads to large deviations from General Relativity in the strong field.  
This class, however, has recently been shown to be in conflict with Solar System observations when accounting for the cosmological evolution of the scalar field.
We study whether this remains the case when the exponent of the conformal factor is positive, as well as in another class of theories defined by a hyperbolic conformal factor.
We find that in both of these scalar-tensor theories, Solar System tests are passed only in a very small subset of parameter space, for a large set of initial conditions compatible with Big Bang Nucleosynthesis. 
However, while we find that it is possible for neutron stars to scalarize, one must carefully select the coupling parameter to do so, and even then, the scalar charge is typically two orders of magnitude smaller than in the negative exponent case.
Our study suggests that future work on scalar-tensor gravity, for example in the context of tests of General Relativity with gravitational waves from neutron star binaries, should be carried out within the positive coupling parameter class.

\end{abstract}

\maketitle
\allowdisplaybreaks[4]

\section{Introduction}
\label{Intro}

	General Relativity (GR) has been shown to be consistent with all current observations, including those in the Solar System (SS)~\cite{2014LRR....17....4W,2003Natur.425..374B}, with binary pulsars~\cite{2012MNRAS.423.3328F}, and recently with gravitational waves through the merging of two black holes detected by advanced LIGO~\cite{Collaboration:2016ki,2016PhRvL.116v1101A}. Scalar-tensor theories (STTs) of gravity are among the most natural extensions to GR~\cite{1970PhRvD...1.3209W,1970ApJ...161.1059N,1968IJTP....1...25B} because they are both well-posed~\cite{Salgado:881405} and well-motivated from string theory~\cite{polchinski1998string,green1988superstring}, quantum field theory~\cite{birrell1984quantum}, and cosmology~\cite{Weinberg:2008hq,faraoni2004cosmology,2012PhR...513....1C}. These theories have seen a recent revival given that they can lead to large deviations from GR in strong-field scenarios, making them a natural theory choice to test GR with gravitational wave observations. 	
	
%
	STTs modify GR by introducing a scalar field that couples to the metric non-minimally, thus forcing matter to respond both to the metric tensor and to the scalar field. The easiest way to see this is to (conformally) transform the metric tensor into the \emph{Einstein frame}, in which the action looks identical to the Einstein-Hilbert one, but with the matter sector responding to the product of the conformally-transformed metric and the conformal factor. Therefore, STTs can be classified by the specific functional form of the conformal factor. The most studied class is that originally proposed by Damour and Esposito-Far\`ese (DEF)~\cite{Damour:1992we,Damour:1993hw}, in which the conformal factor is a Gaussian in the Einstein-frame scalar field with an exponent proportional to a constant $\beta$. Another class that has recently captured some attention is that in which the conformal factor is proportional to a certain power of a hyperbolic cosine with argument proportional to the product of the Einstein-frame scalar field and a constant $\beta$~\cite{Mendes:2016fby,Mendes:2015gx,2011PhRvD..83h1501P,Lima:2010dh}. In this paper, we will study both classes, referring to the former as \emph{exponential} and the latter as \emph{hyperbolic}.
	\begin{figure*}[!t]
		\centering
		\includegraphics[width=7in]{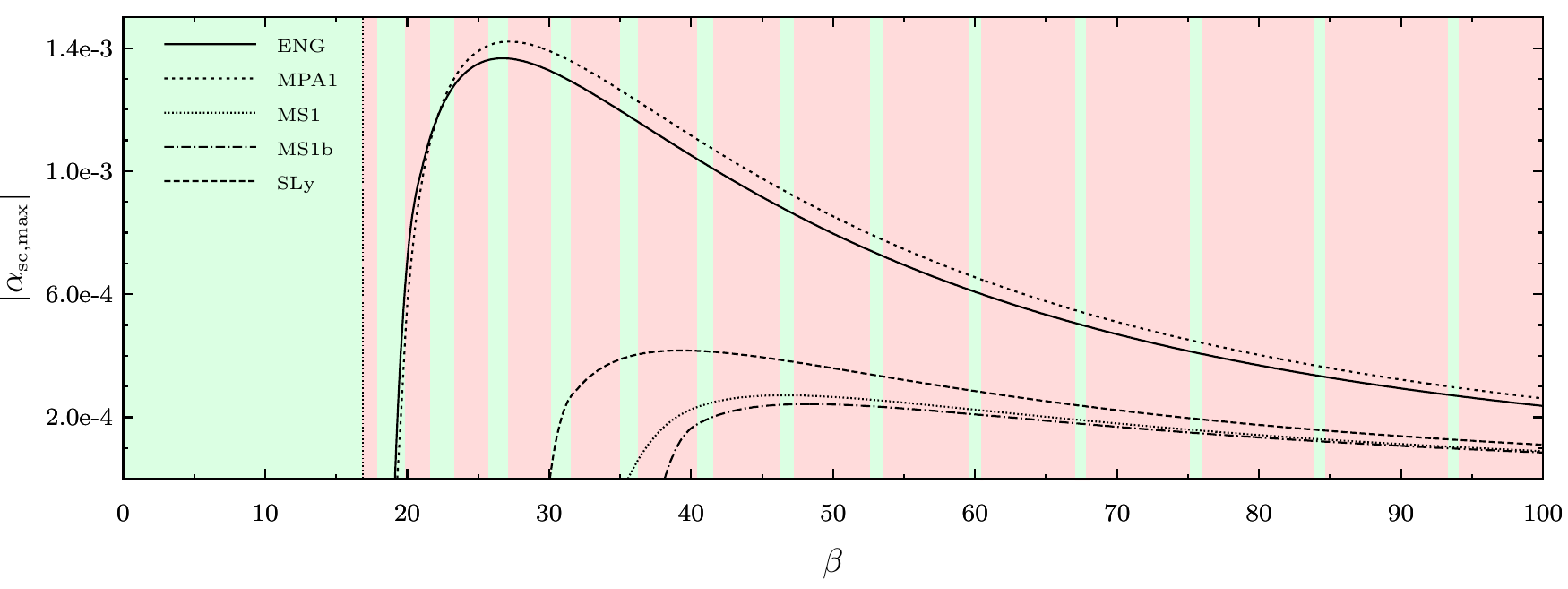}
		\caption{ \label{fig:max_charge_beta} (Color Online) Maximum value of scalar charge $\alpha_\SC$ inside stable NSs as a function of the coupling parameter $\beta$ in hyperbolic STTs with 5 different EoSs. In the background, green (red) regions correspond to values of $\beta$ that are consistent (inconsistent) with Solar System observations after cosmological evolution. Observe that the green regions become thinner and more sparse as $\beta$ increases. The bounds shown here are the most stringent ones one can place with initial conditions that saturate BBN constraints ($\zeta=1$ in Eq.~(\ref{eq:BBN_pot_constraint})). 
		}
	\end{figure*}
	
	Perhaps one of the most interesting consequences of STTs is that neutron stars (NSs) can undergo what is known as \emph{scalarization}~\cite{Damour:1992we,Damour:1993hw}, in which the scalar field is amplified above its background value inside the NS. Three types of scalarization processes have been discovered, which can be classified by the physical process that causes the activation of the scalar field. For isolated NSs, \emph{spontaneous scalarization} occurs once a critical density is reached inside the star~\cite{Damour:1992we,Damour:1993hw,Damour:1996ke,Harada:342978}. In binary systems, NSs can undergo either \emph{dynamical} or \emph{induced} scalarization~\cite{2014PhRvD..89d4024P,2013PhRvD..87h1506B,2014PhRvD..89h4005S,2015PhRvD..91b4033T}. The former occurs when the binding energy of the orbit reaches a critical value, causing the violent activation of the scalar field, while the latter occurs when one member of the binary is already scalarized and it induces a scalar charge in its companion. Until recently, scalarization was thought to only occur in the exponential class of STTs if $\beta<0$, because then a certain scalar instability can occur for most equations of state (EoSs). But recently, a non-vanishing scalar charge has also been shown to occur inside NSs when $\beta>0$ in both exponential and hyperbolic STTs with certain EoSs~\cite{2016PhRvD..93d4009P,Mendes:2016fby,Mendes:2015gx}. In these cases, however, the activation of the scalar charge (for example, as a function of ADM mass) is not nearly as sudden as in the $\beta<0$ case and the maximum scalar charge is typically significantly smaller. 
	
	STTs have also been popular because of the belief that they can be made to pass Solar System constraints by appropriately choosing the asymptotic value of the scalar field, but recently this belief was proven to be incorrect~\cite{Anderson:2016fi,Sampson:2014qqa,PhysRevLett.70.2217,PhysRevD.48.3436}. Solar System observations place a very strict constraint on exponential STTs because the asymptotic value of the scalar field in the Solar System cannot be chosen freely, but rather it must be chosen consistently with the field's prior cosmological evolution. The cosmological evolution of the universe forces the scalar field to grow rapidly with redshift when $\beta < 0$ (for all but very fine-tuned initial conditions), leading to clear violations of Solar System tests today, most notably in the consistency of the Shapiro time delay and the perihelion shift of Mercury with the GR predictions~\cite{Bertotti:2003rm}. A previous study~\cite{Anderson:2016fi} attempted to remedy this problem by modifying the form of the exponential conformal factor by adding a term of higher order in the Einstein frame scalar field. While this was enough to ensure Solar System tests are passed today, it reduced the scalar charge in NSs by orders of magnitude.
	
	\begin{table}[h!]
		\centering
		\begin{tabular*}{3.41in}{|L{0.8 in} |C{0.8 in} C{0.8 in} C{0.77 in}|L{1 pt} }
			\hline
			Theory & SS tests? & $\alpha_{\rm sc,max}$ & Stable NS? &\\ [3 pt]
			\hline 
			$\beta<0$ {\footnotesize exp.} & $\times$ & ${\cal{O}}(10^{-1})$ & \checkmark  &\\[2 pt]
			\hline
			$\beta>0$ {\footnotesize exp.} & \checkmark & ${\cal{O}}(10^{-3})$ & $\times$ &\\[2 pt]
			\hline
			$\beta<0$ {\footnotesize hyp.} & $\times$ & ${\cal{O}}(10^{-1})$ & \checkmark  &\\[2 pt]
			\hline
			$\beta>0$ {\footnotesize hyp.} & \checkmark & ${\cal{O}}(10^{-3})$ & \checkmark &\\ [2 pt]
			\hline
		\end{tabular*}
		\caption{\label{tab:ST_props} A summary of the various properties of two classes of STTs. The columns correspond to the following: Do they pass Solar System tests after cosmological evolution? What is the typical magnitude of the maximum scalar charge they can support inside NSs? Do they allow for stable \emph{and} scalarized NS solutions? }
	\end{table}
	%
	%
	
	Given all of this, it is then natural to explore exponential and hyperbolic STTs with $\beta>0$ in more detail to determine if Solar System constraints can be placed, and if so, the degree to which NSs continue to scalarize within the allowed region of parameter space. We first study Solar System constraints by exploring the cosmological behavior of STTs from the time of Big-Bang Nucleosynthesis (BBN) until today. For any particular value of $\beta$ and any particular set of initial conditions consistent with BBN constraints, the value of the scalar field today will either be consistent with Solar System observations or it will violate them, therefore determining if that value of $\beta$ is part of the viable region of parameter space for those initial conditions. 
	
	Figure~\ref{fig:max_charge_beta} shows the $\beta$ (green) regions that allow hyperbolic STTs to pass Solar System tests when choosing initial conditions that saturate BBN constraints. For all viable initial conditions (including those that saturate BBN constraints), hyperbolic STTs pass Solar System tests when $0<\beta \lesssim 17$, while exponential STTs pass when $0<\beta \lesssim 24$ (not shown in the figure). Moreover, there are special values of the initial conditions for which both theories pass Solar System tests above these critical $\beta$ values, which are shown as thin green regions in Fig.~\ref{fig:max_charge_beta}. This is because the cosmological evolution of the scalar field has damped oscillations with redshift, thus allowing for the possibility that today the field happens to be in a state in which Solar System tests are passed, although this may not be the case in the future. 
	
	The constraints reported above can be relaxed by fine-tuning the initial conditions at the time of BBN. Cosmological observations and the theory nucleosynthesis require that the scalar field at the time of BBN be in a small region near the minimum of an effective potential. Saturating current constraints on the speed-up factor put the scalar as far away from the minimum as possible. In principle, however, one can relax this assumption and place the scalar closer to the minimum, making it easier for the theory to satisfy current Solar System constraints. A random distribution of viable initial conditions relaxes the bounds reported above to $0 < \beta \lesssim 34$ and $0 < \beta \lesssim 25$ in the exponential and hyperbolic cases respectively. Larger values of $\beta$ would typically require fine-tuning of initial conditions for Solar System tests to be passed.

%
	 
	With this at hand, we then explore whether spontaneous scalarization still occurs in exponential and hyperbolic STTs within the viable $\beta$ regimes of parameter space. In the exponential STT case, scalarized NSs have already been shown to be unstable to gravitational collapse for such values of $\beta$~\cite{Mendes:2016fby}. In the hyperbolic case, however, scalarized NSs can be stable for certain EoSs, but the scalar charge is typically two orders of magnitude smaller than typical charges in the $\beta<0$ exponential STT case.  Figure~\ref{fig:max_charge_beta} plots the maximum scalar charge for stable NSs in hyperbolic STTs for a large range of $\beta$ using different EoSs. None of the EoSs that we consider here give stable scalarized NS solutions for $\beta \lesssim 17$, but they do allow for stable scalarized stars when $\beta \gtrsim 20$. Moreover, there exist (small) periodic regions of viable parameter space that allow for scalarized NSs (green regions that have non-zero scalar charge in them), although these separate and become thinner as $\beta$ increases. These general conclusions are summarized in Table~\ref{tab:ST_props}. 
	
	Our results are directly relevant to studies that constrain deviations from GR in the strong field, for example with gravitational waves. Most of the latter have so far focused on the massless exponential class of STTs with $\beta < 0$, but these have already been shown to be incompatible with Solar System observations~\cite{Sampson:2014qqa,Anderson:2016fi}. We will here show that the massless exponential class of STTs with $\beta>0$ does pass Solar System tests for a range of $\beta$, but these theories have already been shown to disallow for stable scalarized NSs~\cite{Mendes:2016fby}. We will also show that the hyperbolic class of STTs with $\beta>0$ also passes Solar System tests for a range of $\beta$, and these theories do allow for stable scalarized stars~\cite{Mendes:2016fby} making them much more appropriate for gravitational wave studies. Our results, however, also indicate that the amount of scalar charge in such theories is drastically smaller than what is obtained in the massless exponential case with $\beta < 0$. This raises the question of whether such small charges can be constrained in practice with second-generation ground-based gravitational wave detectors. A more detailed analysis is required to address this last question.
		
	The rest of this paper will address the points above in more detail. We first lay out our notation and the background of STTs in Sec.~\ref{The Basics of Scalar-Tensor Theories}. We next go into details of the cosmological evolution of the scalar field in both theories in Sec.~\ref{Cosmological Evolution and solar system Constraints} and place constraints on $\beta$ from Solar System observations. Section~\ref{Neutron Stars and Scalarization} addresses NS solutions in detail, first for $\beta<0$ to build a foundation and then for $\beta>0$. Finally, we end with our conclusions in Sec.~\ref{Conclusion} and discuss the implications of our results. We adopt units in which $c=1$ but restore cgs units in our discussion of NSs when appropriate.

\section{The Basics of Scalar-Tensor Theories}
\label{The Basics of Scalar-Tensor Theories}

	In this section we present the details of the class of theories we investigate and establish notation, following mostly the presentation in~\cite{Anderson:2016fi}. The STTs we consider in this paper can be defined by the action $S_\J = S_{\J,g} + S_{\J,\mat}$, where the gravitational part is given by
	\be
		  S_{\J,g} \eq \int d^4 x \dfrac{\sqrt{-g}}{2\kappa}\left[\phi R - \dfrac{\omega(\phi)}{\phi}\partial_\mu \phi \, \partial^\nu \phi\right]\,\,,
	\label{eq:action-jordan}
	\ee
	and where $g$ and $R$ are the determinant and Ricci scalar associated with the Jordan-frame metric $g_{\mu\nu}$, $\omega(\phi)$ is a coupling function for the scalar field $\phi$, and $\kappa=8 \pi G$ with $G$ the bare gravitational constant. The matter action $S_{\J,\mat}[\chi,g_{\mu\nu}]$ is a functional of the matter fields $\chi$, which couple directly to the Jordan frame metric $g_{\mu\nu}$, meaning that STTs of this form are metric theories of gravity.
	
	One can take the action in Eq.~(\ref{eq:action-jordan}) and write it in a form resembling the Einstein-Hilbert action in GR through a conformal transformation $g_{\mu\nu} = A^2(\varphi)g_{\mu\nu}^*$ where we refer to $g_{\mu\nu}^*$ as the Einstein-frame metric. By choosing the conformal factor $A(\varphi)$ such that
	\be
		A^2(\varphi) \eq \phi^{-1}\,\,,
	\label{eq:def-conformal_factor}
	\ee
	and providing an explicit relationship between $\varphi$ and $\phi$ via
	\be
		\alpha^2(\varphi) \eq \left(\dfrac{d \ln A(\varphi)}{d \varphi}\right)^2 \eq \dfrac{1}{3 + 2 \omega(\phi)}\,\,,
	\label{eq:def-omega}
	\ee%
	the Einstein-frame action becomes
	\be
	\begin{split}
		S_{\E} \eq &\int d^4x \dfrac{\sqrt{-g_*}}{2\kappa}\Big[R_* - 2g_*^{\mu\nu}\partial_\mu\varphi\,\partial_\nu\varphi\Big] \\
		&+ S_{\E,\mat}\big[\chi,A^2(\varphi)g_{\mu\nu}^*\big]\,\,,
	\end{split}
	\label{eq:part1:r_cross_Q}
	\ee
	where $g_*$ and $R_*$ are the determinant and Ricci scalar associated with the Einstein-frame metric $g_{\mu\nu}^*$. Notice that the matter degrees of freedom $\chi$ now couple to the product of the conformal factor $A(\varphi)$ and the Einstein-frame metric. For later convenience, we define a conformal coupling potential via $\alpha = \partial V_{\alpha}/\partial \varphi$, such that
	\be
		V_{\alpha} (\varphi) \eq \ln A(\varphi) \,\,,
	\label{eq:conformal_potential}
	\ee
	and note that the curvature of this potential is then given by
	\be
		\beta(\varphi) \eq \dfrac{\partial^2 V_\alpha}{\partial\varphi^2} \eq \dfrac{\partial\alpha(\varphi)}{\partial\varphi}\,\,.
	\label{eq:conformal curvature}
	\ee
	The potential in Eq.~(\ref{eq:conformal_potential}) will later play the role of an effective potential that the scalar field evolves in cosmologically. For the rest of this paper we will strictly refer to $A(\varphi)$ as the conformal factor, $\alpha(\varphi)$ as the (conformal) coupling, and $V_\alpha(\varphi)$ as the (conformal) coupling potential.
	
	Variation of the Einstein-frame action yields the field equations
	\ba
		R_{\mu\nu}^* &\eq& \kappa\left(T_{\mu\nu}^* - \dfrac{1}{2}g_{\mu\nu}^* T^*\right)\,\,,
	\label{eq:EEs}
	\\
		&\Box_*\varphi& \eq -\dfrac{\kappa}{2}\alpha(\varphi)T^{\mat,*}\,\,,
	\label{eq:KG}
	\ea
	where $\Box_*$ is the Einstein-frame covariant wave operator and $T^{\mat,*}$ is the trace of the Einstein-frame matter stress-energy tensor (SET) defined by
	\be
		T_{\mu\nu}^{\mat,*} \equiv \dfrac{2}{\sqrt{-g_*}} \dfrac{\delta S_{\E,\mat}\big[\chi,A^2(\varphi) g_{\mu\nu}^*]}{\delta g_{\mu\nu}^*}\,\,.
	\label{eq:def-SETmat}
	\ee
	One can see that the Einstein-frame stress-energy tensor is related to the Jordan-frame one via $T^{\mu\nu}_{\mat,*} = A^6 T^{\mu\nu}_{\mat}$ by applying the conformal transformation to Eq.~(\ref{eq:def-SETmat}). The field equations also depend on the total stress-energy tensor, which is the sum of the matter and scalar field stress-energy tensors 
	\be
		T_{\mu\nu}^* \eq T_{\mu\nu}^{\mat,*} + T_{\mu\nu}^{\varphi,*}\,\,,
	\label{eq:def-SETtotal}
	\ee
	where
	\be
		\kappa T_{\mu\nu}^{\varphi,*} \eq 2(\partial_\mu \varphi)(\partial_\nu \varphi) - g_{\mu\nu}^*(\partial_\sigma \varphi)(\partial^\sigma \varphi)\,\,.
	\label{eq:def-SETscalar}
	\ee

	Our entire study, both cosmologically and for NSs, adopts a perfect fluid representation of the matter in the Jordan frame. However, this is also true in the Einstein frame since they are related through a conformal transformation, and thus we use
	\be
		T_{\mu\nu}^{\mat,*} \eq (\epsilon_* + p_*)u_\mu^* u_\nu^* + p_* g_{\mu\nu}^*\,\,,
	\label{eq:def-perfect_fluid}
	\ee
	where $\epsilon_*$ is the energy density of the fluid, $p_*$ is the pressure, and $u^*$ is the 4-velocity, all in the Einstein frame. Normalization of the 4-velocity $g_{\mu\nu}^* u_*^\mu u_*^\nu = -1 = g_{\mu\nu} u^\mu u^\nu$ leads to the relationship $u_*^\mu = A u^\mu$. Using this result in conjunction with $T_*^{\mu\nu} = A^6 T^{\mu\nu}$ allows one to derive the direct relations $\epsilon_* = A^4 \epsilon$ and $p_* = A^4 p$ between the Einstein and Jordan frame density and pressure.
	
	The choice of $A(\varphi)$, or consequently $\alpha(\varphi)$, defines a particular theory and therefore plays a critical role in testing that theory with observations. Specifically, these functions determine the local value of Newton's gravitational constant~\cite{1970ApJ...161.1059N}
	\be
		G_N \eq G \big[A^2\,(1 + \alpha^2)\big]_{\varphi_0}\,\,,
	\label{eq:GN}
	\ee
	where $\varphi_0$ is the value of the scalar field today determined cosmologically. They also determine the local value of the parameterized post-Newtonian (PPN) parameters~\cite{1972ApJ...177..757W,1972ApJ...177..775N}. The $\gamma_\ppN$ parameter, which measures the spatial curvature due to a unit rest mass~\cite{TEGP,2014LRR....17....4W}, is given in massless STTs by
	\be
		|1-\gamma_{\ppN}|_{\varphi_0} \eq \frac{2 \alpha_0^{2}}{1 + \alpha_0^{2}}\,\,,
	\label{eq:gamma_ppn}
	\ee
	where $\alpha_0$ is the conformal coupling evaluated at the present day value of the scalar field $\varphi_0$. There exist another such expression for the $\beta_\ppN$ parameter which is given by 
	\be
		|1-\beta_\ppN|_{\varphi_0} \eq \dfrac{1}{2}\dfrac{\beta_0 \alpha_0^2}{(1+\alpha_0^2)^2}\,\,,
	\label{eq:beta_ppn}
	\ee
	where $\beta_0 = \partial \alpha/\partial\varphi|_{\varphi_0}$. The constraints $|1-\gamma_\ppN| \lesssim 2.3\times 10^{-5}$ and $|1-\beta_\ppN| \lesssim 8\times 10^{-5}$ ~\cite{2014LRR....17....4W}, from the verification of the Shapiro time delay by the Cassini spacecraft and the perihelion shift of Mercury respectively ~\cite{TEGP}, provide the weak field constraints on the theories we consider in this paper.
	
\section{Cosmological Evolution and Solar-System Constraints}
\label{Cosmological Evolution and solar system Constraints}

	To determine if a particular theory is consistent with Solar System tests, we must first understand the cosmological evolution of the scalar field. We here review the cosmological evolution equations and continue to establish notation, following again mostly the presentation in~\cite{Anderson:2016fi}. We adopt a spatially flat Friedmann-Roberston-Walker (FRW) metric in the Einstein-frame
	\be
		ds^2_* = -dt^2_* + a_*^2(dr^2_* + r^2 d\Omega^2_*)\,\,,
	\label{eq:frw_metric}
	\ee
	where $a_*$ is the Einstein-frame scale factor. The Einstein-frame field equations then become
	\begin{align}
		3 H_{*}^{2} &= \kappa \epsilon_* + \dot{\varphi}^{2} \,\,,
	\label{eq:friedmann1}
		\\
		-3 \frac{\ddot{a}_{*}}{a_{*}} &= \frac{\kappa}{2} \epsilon_* \left(1 + 3 \lambda\right) + 2 \dot{\varphi}^{2}\,\,,
	\label{eq:freidmann2}
		\\
		\ddot{\varphi} + 3 H_{*} \dot{\varphi} &= -\dfrac{\kappa}{2} \alpha \epsilon_* (1 - 3 \lambda) \,\,,
	\label{eq:cosmo_KG}
	\end{align}
	where overhead dots stand for derivatives with respect to Einstein-frame coordinate time $t_*$, $H_* = \dot{a}_*/a_*$ is the Einstein-frame Hubble parameter, $\epsilon_*$ and $p_*$ are the energy density and pressure of all components of the universe (matter, radiation, dark energy), and $\lambda = p_*/\epsilon_*$ is the usual cosmological EoS parameter. The Einstein-frame variables can be transformed to Jordan-frames variables via the conformal transformation mentioned earlier: $g_{\mu\nu} = A^2(\varphi) g_{\mu\nu}^*$ which yields $dt = A(\varphi)dt_*$ and $a = A(\varphi)a_*$. Following Ref.~\cite{Anderson:2016fi}, we assume that any pressure and energy density associated with $\varphi$ will be negligible compared to other energy components of the universe. 
	
	Equations~(\ref{eq:friedmann1})--(\ref{eq:cosmo_KG}) do not lend themselves to a simple analytic solution for the evolution of the scalar field $\varphi$. However, by defining a new time coordinate $d\tau = H_* dt_*$, one can decouple the evolution equations. Under this time transformation one finds that the time derivatives become
	\ba
		\dot{\varphi} &=& H_* \varphi'\,\,,
		\\
		\ddot{\varphi} &=& H_*^2\varphi'' +\dot{H}_*\varphi'\,\,,
		\\
		\dfrac{\ddot{a}}{a} &=& \dot{H}_* + H_*^2\,\,,
	\ea
	where primes denote derivative with respect to $\tau$. Using these redefinitions, Eqs.~(\ref{eq:friedmann1})--(\ref{eq:cosmo_KG}) can be simplified into a single equation for the evolution of the scalar field
	\be
		\dfrac{2}{3 - {\varphi'}^2}\varphi'' \epsilon_* + \epsilon_*(1-\lambda)\varphi' = -\alpha(\varphi)\epsilon_*(1 - 3\lambda)\,\,.
	\label{eq:scalar_evolution1}
	\ee
	It is important to note here that if one wishes to consider more than a single component universe, $\epsilon_*$ must not be divided out of the equation. In general, each term containing $\epsilon_*$ and $\lambda$ will be a sum over all components of the universe, and therefore, $\epsilon_{*}$ cannot simply be removed from the equation. 
	
	To have a complete description of Eq.~(\ref{eq:scalar_evolution1}) one must understand its evolution in $\tau$. From conservation of energy in the Jordan-frame one finds
	\be
		d(\epsilon a^3) = -p(\epsilon)d(a^3)\,\,,
	\label{eq:SET_conservation_time}
	\ee
	which, using $\lambda = p/\epsilon$, can be written as
	\be
		\epsilon = a^{-3(1+\lambda)}\epsilon_0\,\,,
	\label{eq:density_Jframe}
	\ee
	where $\epsilon_0$ is the value of the Jordan-frame energy density measured today (i.e. when $a=1$). To transform Eq.~(\ref{eq:density_Jframe}) to the Einstein frame we make use of the fact that $d\tau = H_* dt_*$ to write $a_* = e^{\tau}$ and then use $a = A(\varphi) a_*$ to write
	\be
		a = A(\varphi)e^{\tau}\,\,.
	\label{eq:rho_jordan}
	\ee
	We can then use $\epsilon_* = A^4 \epsilon$ in combination with Eq.~(\ref{eq:rho_jordan}) to write the Einstein-frame energy density as
	\be
		\epsilon_* = \epsilon_0 e^{-3(1+\lambda)\tau}A^{1-3\lambda}\,\,.
	\label{eq:rho_einstein}
	\ee
	If we divide Eq.~(\ref{eq:scalar_evolution1}) by the critical density $\epsilon_{\text{crit}}$ and use the result of Eq.~(\ref{eq:rho_einstein}) the scalar field evolution equation becomes
	\be
		\dfrac{2\varphi''}{3 - {\varphi'}^2} C_1(\varphi,\tau) + \varphi' C_2(\varphi,\tau) = -\alpha(\varphi) C_3(\varphi,\tau)\,\,,
	\label{eq:scalar_evo}
	\ee 
	where we have defined
	\ba
		C_1(\varphi,\tau) &=& \sum\limits_i \Omega_{i,0}\,e^{-3(1+\lambda_i)\tau}\,A^{(1-3\lambda_i)}\,\,,
		\label{eq:c1}
		\\
		C_2(\varphi,\tau) &=& \sum\limits_i \Omega_{i,0}(1-\lambda_i)\,e^{-3(1+\lambda_i)\tau}\,A^{(1-3\lambda_i)}\,\,,
		\label{eq:c2}
		\\
		C_3(\varphi,\tau) &=& \sum\limits_i \Omega_{i,0}(1-3\lambda_i)\,e^{-3(1+\lambda_i)\tau}\,A^{(1-3\lambda_i)}\,\,,
		\label{eq:c3}
	\ea
	where $\Omega_{i,0} = \epsilon_{i,0}/\epsilon_{\text{crit}}$ is the standard density parameter as seen today for the $i$th energy component.
	
	The evolution described in Eq.~(\ref{eq:scalar_evo}) is reminiscent of a damped oscillator. Assuming the density functions $C_1$, $C_2$, and $C_3$ only contain a single energy component for simplicity, it becomes apparent that there is a velocity dependent mass appearing in front of the second derivative, a damping term proportional to the velocity, and on the right-hand side of the equations there is a potential gradient influencing the overall evolution of the scalar field, as first pointed out in~\cite{PhysRevLett.70.2217,PhysRevD.48.3436}. Even if the density functions are not constant in $\tau$ or $\varphi$, the damped oscillator picture still holds, as we will show numerically later in this section, and it will thus provide a useful analogy to extract a physical understanding from our numerical calculations.

\subsection{Two ST models}

	Thus far we have remained model independent and have not made any assumptions about the form of $\alpha(\varphi)$. In this section we define the two STTs that we study in this paper. The first model we consider is the standard DEF theory, defined by
	\ba
		A(\varphi) &=& e^{\frac{1}{2}\beta \varphi^2}\,\,,
		\label{eq:A_DEF}
		\\
		V_{\alpha}(\varphi) &=& \dfrac{1}{2}\beta \varphi^2\,\,,
		\label{eq:V_DEF}
		\\
		\alpha(\varphi) &=& \beta \varphi\,\,,
		\label{eq:alphaDEF}\\
		\beta(\varphi) &=& \beta\,\,,
		\label{eq:betaDEF}
	\ea
	which we will refer to as the \emph{exponential model}. The other model we consider is that introduced by Mendes~\cite{Mendes:2016fby,Mendes:2015gx}, defined by
	\ba
		A(\varphi) &=& \left[\cosh(\sqrt{3}\beta\varphi)\right]^{\frac{1}{3 \beta}}\,\,,
		\label{eq:A_COSH}
		\\
		V_{\alpha}(\varphi) &=& \dfrac{1}{3 \beta}\ln{\left[\cosh(\sqrt{3}\beta\varphi)\right]}\,\,,
		\label{eq:V_COSH}\\
		\alpha(\varphi) &=& \dfrac{\tanh(\sqrt{3}\beta \varphi)}{\sqrt{3}}\,\,,
		\label{eq:alphaCOSH}\\
		\beta(\varphi) &=& \beta \,\text{sech}^2(\sqrt{3}\beta\varphi)\,\,,
		\label{eq:betaCOSH}
	\ea
	which we will refer to as the \emph{hyperbolic model}. Notice that $\beta$ enters both models as a free parameter that will quantify the degree of departure from GR (with $\beta = 0$ reducing the theory to GR). The conformal factor here is motivated as an analytic approximation to the conformal factor one would find if the action also included terms proportional to $R \; \varphi^{2}$ in the Einstein frame~\cite{Mendes:2016fby,2011PhRvD..83h1501P,Lima:2010dh,Mendes:2015gx}, motivated by quantum field theory considerations~\cite{birrell1984quantum}.
	\begin{figure*}[t]
		\centering
		\includegraphics[width=3.5in]{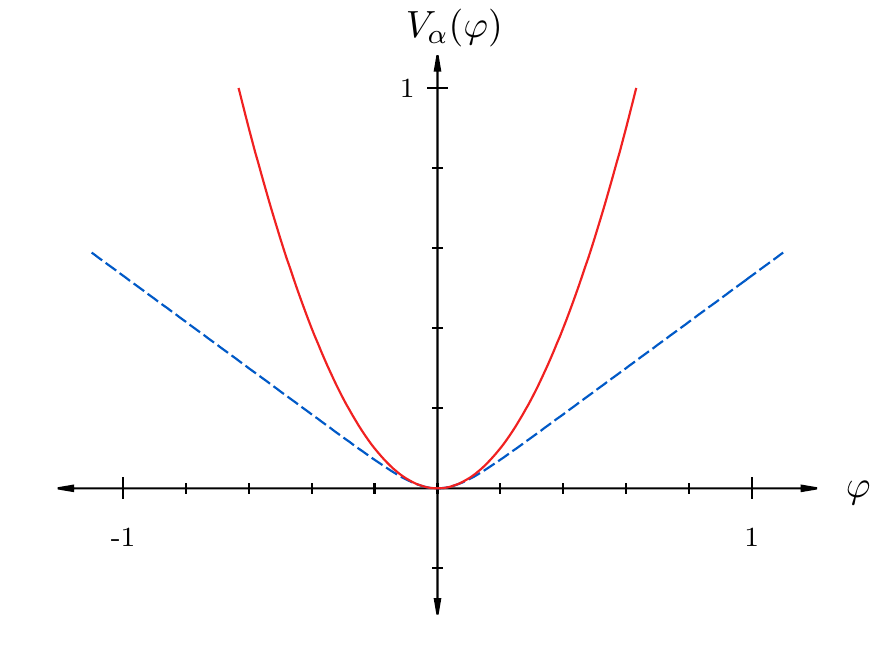}
		\includegraphics[width=3.5in]{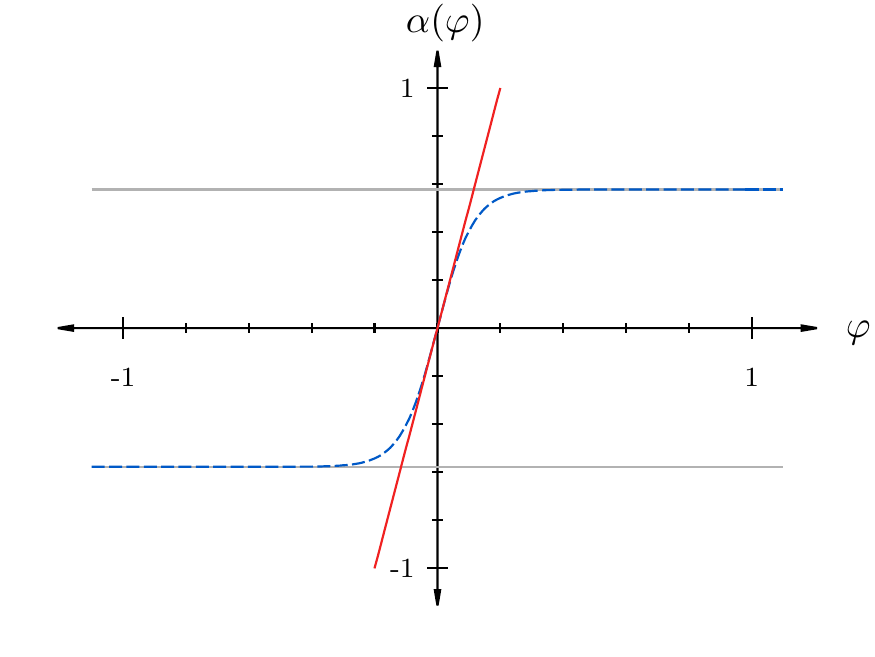}
		\caption{ \label{fig:potentials_couplings} (Color Online) Schematic diagrams for both the conformal coupling potential $V_\alpha$ (left) and the conformal coupling $\alpha(\varphi)$ (right). The exponential STT is represented by solid red curves and the hyperbolic theory by dashed blue curves. The horizontal gray lines at $\pm 1/\sqrt{3}$ in the right panel correspond to the limiting values of $\alpha(\varphi)$ in the hyperbolic theory. These figures were constructed with $\beta=5$, but the qualitative features remain the same for all values of $\beta$.
		}
	\end{figure*}
	
	In the context of an oscillator, $V_{\alpha}$ is the potential that sources the scalar field evolution through its gradient $\alpha(\varphi)$. In the case of exponential STTs, this is simply a parabola, cf. the left panel of Fig.~\ref{fig:potentials_couplings}, and therefore the scalar field evolves analogously to a damped harmonic oscillator in $\tau$. In the case of hyperbolic coupling, however, that potential is no longer a simple parabola and while we expect the evolution of the scalar field to still be oscillatory, it should be qualitatively different than in the exponential STT case. The key difference between the potentials in these theories lies in their behavior away from their minima ($\varphi=0$). In the exponential case, the gradient of the potential grows linearly as we increase $\varphi$ (or $\beta$ for that matter) where for hyperbolic coupling the gradient becomes constant; both behaviors can be seen in the right panel of Fig.~\ref{fig:potentials_couplings}. The behavior of these potentials away from their minima plays a key role in the behavior of the scalar field, both in cosmology and in NSs, as we will show in the following sections.

\subsection{BBN Constraints and Initial Conditions}

	To determine the evolution of the scalar field one must first determine a set of initial conditions that it must obey at some point in the past. The natural first choice is to determine initial conditions at the very beginning of the universe, or at the very least at the end of inflation. However, there is substantial uncertainty in the true physics of the early universe, and thus, any constraints would be subject to this uncertainty. Following the work in~\cite{Anderson:2016fi}, we will use the observational constraints from Big-Bang-Nucleosynthesis (BBN) to determine the behavior of the scalar field at matter-radiation equality ($Z \sim 3600$), which we will use as our ``initial'' point in time for our numerical integrations.

	Consider an era in which the universe is radiation dominated, as was the case in the early universe after inflation, when $a \sim 10^{-5}$. In this case we can neglect the energy density contributions from matter and dark energy and write Eqs.~(\ref{eq:scalar_evo})--(\ref{eq:c3}) as simply
	\be
		\dfrac{2}{3 - {\varphi'}^2}\varphi'' + \dfrac{2}{3}\varphi' = 0\,\,,
	\label{eq:evo_radition}
	\ee
	where we have divided out the common energy density terms. Notice that the coupling function $\alpha(\varphi)$ does not appear in this equation, meaning that the scalar field evolution during a purely radiation dominated universe will always be the same, regardless of the coupling function of the theory. Equation~(\ref{eq:evo_radition}) has a solution of the form (following the notation in Refs.~\cite{PhysRevLett.70.2217,PhysRevD.48.3436})
	\be
		\varphi(\tau) = \varphi_{r} - \sqrt{3}\ln\left[K e^{-\tau} + (1+ K^2 e^{-2\tau})^{1/2}\right]\,\,,
	\label{eq:scalar_radiation}
	\ee
	where
	\be
		K = \dfrac{\varphi_i' \sqrt{3}}{\sqrt{1 - \varphi_i'^2/3}}\,\,,
	\ee
	with $\varphi_r$ a constant and $\varphi_i'$ the scalar velocity upon exiting inflation. The only necessary assumption that we must make here is that the scalar velocity not be very close to its limiting value of $\sqrt{3}$. This assumption is valid because if the scalar velocity ever does reach $\sqrt{3}$ then all dynamics of the evolution are removed, which can be seen from Eq.~(\ref{eq:scalar_evo}). When $\varphi'=\sqrt{3}$ the only solution to Eq.~(\ref{eq:scalar_evo}) is one in which the field grows linearly in time, which certainly violates Solar System constraints today since the right-hand side of Eq.~(\ref{eq:gamma_ppn}) would asymptote to unity. Thus, we can conclude that $\varphi'<\sqrt{3}$ for \emph{all times} in the past and from Eq.~(\ref{eq:scalar_radiation}) we can conclude that $\varphi'(\tau)$ is exponentially damped to zero during radiation domination. 

	To constrain the initial value of the scalar field we use BBN constraints on the gravitational constant. The speed-up factor $\xi_{\bbn} := H/H_{GR}$ quantifies any differences between the actual Hubble parameter $H$ and the the Hubble parameter predicted by GR, $H_{GR}$; these differences are caused by deviations from the standard gravitational constant. The observed Hubble parameter can be derived from Eq.~(\ref{eq:friedmann1}) with $\dot{\varphi}=0$ (as we have just argued to be true during the radiation era), and is given by $H=(8\pi/3)GA^2_{\mbox{\tiny $R$}}\epsilon_{\mbox{\tiny $R$}}$ where $\epsilon_{\mbox{\tiny $R$}}$ and $A_{\mbox{\tiny $R$}}$ are the Jordan-frame energy density and conformal factor, respectively, evaluated at the time of BBN. The expansion rate predicted in GR takes the form $H_{GR}^2 = (8\pi/3)G_N\epsilon_{\mbox{\tiny $R$}}$, with $G_N$ given in Eq.~(\ref{eq:GN}). The speed-up factor then becomes
	\be
	\xi_{\bbn} = \left(\dfrac{H}{H_{GR}}\right) = \left(\dfrac{G A_{\mbox{\tiny $R$}}^2}{G_N}\right)^{1/2} = \dfrac{1}{\sqrt{1+\alpha_0^2}}\dfrac{A_{\mbox{\tiny $R$}}}{A_0}\,\,,
	\label{eq:bbn_condition}
	\ee
	where the $R$ and $0$ subscripts indicate values at the end of the radiation era (or at the time of BBN more specifically) and today respectively. 

	Observational bounds on the abundances of Helium place constraints on the expansion rate of the universe \cite{2011LRR....14....2U}, which then implies that 
	\be
		|1 - \xi_{\bbn}| \leq \frac{1}{8}\,\,.
	\label{eq:bbn_constraint}
	\ee
	On the other hand, Solar System tests restrict $\alpha_0^2 \sim 10^{-5}$ which means that to a good approximation we can neglect its contributions in Eq.~(\ref{eq:bbn_condition}) and write $\xi_{\bbn}\sim A_{\mbox{\tiny $R$}}/A_0$. A constraint on $A_{\mbox{\tiny $R$}}$ can be placed through Eq.~(\ref{eq:bbn_constraint}), giving
	\be
		\left|1 - \dfrac{A_{\mbox{\tiny $R$}}}{A_0}\right| \leq \frac{1}{8}\,\,.
	\label{bbn_constraint_ratio}
	\ee
	Saturating this constraint gives $A_{{\mbox{\tiny $R$}}}/A_{0} = 7/8$ or $A_{{\mbox{\tiny $R$}}}/A_{0} = 9/8$. 
	
	From here the particle analogy can be used to understand what exactly this constraint tells us. Relating the constraint on the conformal factor $A$ to the conformal coupling potential $V_{\alpha}$ via Eq.~(\ref{eq:conformal_potential}) one finds
	\be
		V_{\alpha,0} + \ln(7/8) \,\leq\, V_{\alpha,{\mbox{\tiny $R$}}} \,\leq\, V_{\alpha,0} + \ln(9/8)\,\,.
	\label{potential_rad}
	\ee
	This tells us that the scalar field can, at most, be $\ln(9/8)$ higher in the potential than where it is today or $\ln(7/8)$ lower. However, both conformal potentials we consider here, Eqs.~(\ref{eq:V_DEF}) and (\ref{eq:V_COSH}), have a global minimum which the scalar field \emph{must} be near to today in order to satisfy the Cassini bound. Such a condition means the the scalar field cannot be lower in the potential than it is today and therefore only $V_{\alpha,{\mbox{\tiny $R$}}} \,\leq\,  \ln(9/8)$ is a valid constraint (because $V_{\alpha,0}\approx 0)$. For our calculations we will use
	\be
		V_{\alpha,{\mbox{\tiny $R$}}} = \zeta\ln(9/8)\,\,,
	\label{eq:BBN_pot_constraint}
	\ee
	where $\zeta$ is a parameter that ranges from 0 to 1 and will scale the BBN constraint to allow us to sample the full range of possible initial conditions.
	
	Now we must apply the BBN constraint in Eq.~(\ref{eq:BBN_pot_constraint}) to the conformal coupling potentials we consider in this paper. By equating Eq.~(\ref{eq:V_DEF}) and Eq.~(\ref{eq:BBN_pot_constraint}) we find the initial conditions for the exponential STT to be
	\ba
		\varphi_{\mbox{\tiny $R$},\DEF} &=& \sqrt{\dfrac{2}{\beta}\,\zeta\,\ln\left(\dfrac{9}{8}\right)}\,\,,
		\label{eq:DEF_ICs1}
		\\
		\varphi_{\mbox{\tiny $R$},\DEF}'&=& 0\,\,.
	\label{eq:DEF_ICs2}
	\ea
	Likewise, in the hyperbolic case we have
	\ba
	\varphi_{\mbox{\tiny $R$},\COSH} &=& \dfrac{1}{\sqrt{3}\beta}\cosh^{\mbox{\tiny  -1}}\left[\left(\dfrac{9}{8}\right)^{3\beta\zeta}\right]\,\,,
	\label{eq:COSH_ICs1}
	\\ 
	\varphi_{\mbox{\tiny $R$},\COSH}'&=& 0\,\,.
	\label{eq:COSH_ICs2}
	\ea
	%
	%

	The constraints on the conformal factor from BBN also provide information on how far in the past BBN occurred in $\tau$-time. Consider the definition $d\tau = H_* dt_*$ and integrate it to get $\tau = \ln a_*$. Transformation back to the Jordan-frame scale factor via $a = A(\varphi) a_*$ gives
	\be
		\tau = \ln\left(\dfrac{1}{1+Z}\right) - \ln(A)\,\,,
	\label{eq:tau}
	\ee
	where $Z$ is the Jordan-frame redshift related to the Jordan-frame scale factor through $a = 1/(1+Z)$. Because we set $a=1$ today, the value of $\tau$ today must be zero, i.e. $\tau_0 = 0$ . Finding the difference in $\tau$-time between now and when BBN occurred is simply $\tau_0 - \tau_{\mbox{\tiny $R$}}$, or
	\be
		\tau_{\mbox{\tiny $R$}} = \ln\left(\dfrac{1}{1+Z_{\mbox{\tiny $R$}}}\right) - \zeta\,\ln\left(\dfrac{A_{\mbox{\tiny $R$}}}{A_0}\right)\,\,,
	\label{eq:tau_BBN}
	\ee
	where we have made use of the fact that the redshift today is zero. Notice that the ratio $A_{\mbox{\tiny $R$}}/A_0$ is the same ratio that is constrained from BBN, and thus $A_{\mbox{\tiny $R$}}/A_0 \leq 9/8$, which is valid for all values of $\beta$. Therefore, using $Z_{\mbox{\tiny $R$}} = 3600$ and $\zeta=1$, one finds that BBN occurred at most $\tau = \tau_{\mbox{\tiny $R$}} \sim -8.306$. The value of $\tau_{\mbox{\tiny $R$}}$ determined in Eq.~(\ref{eq:tau_BBN}) with $Z_{\mbox{\tiny $R$}} = 3600$ and a given $\zeta$ will mark the initial time of our numerical calculations.

\subsection{ Exponential Theory}

We study the cosmological evolution of the scalar field in detail and place bounds on $\beta$. We start by considering the $\zeta=1$ case, i.e.~saturating the BBN constraints on the speed-up factor, to understand the nature of the evolution for $1<\beta<1000$. From there, we relax this assumption on $\zeta$ and allow $0<\zeta<1$, and we use these results to place bounds on $\beta$ by avoiding fine-tuning scenarios. In order to avoid confusion, we point out that, in general, Solar System constraints on $\gamma_\ppN$ and $\beta_\ppN$ place relatively independent constraints on the value of $\beta$. For completeness, we present both results but in the end our conclusions will be based off of the PPN parameter that places the tightest constraints on $\beta$; in every case that we study, we find that $\beta_\ppN$ consistently places the tightest constraints.\footnote{Technically, constraints on $\dot{G}_N/G_N$, as appearing in Eq.~(\ref{eq:GN}), can also constrain these theories because the oscillating scalar field induces an oscillating value of $G_N$. However, we find that these constraints are insignificant when compared to $\gamma_\ppN$ and $\beta_\ppN$ in the range of parameter space that we explore.}

\subsubsection{Scalar-Field Evolution}
	We solve Eq.~(\ref{eq:scalar_evo}) numerically from the end of the radiation dominated era ($Z \sim 3600$) to the present day. Describing the energy density of the universe as a piecewise function in $\tau$-time would only require one to consider a single energy component of the universe at a time and thus only keep a single term in Eqs.~(\ref{eq:c1})--(\ref{eq:c3})(this leads to a rather simple analytic solution for the scalar field~\cite{Anderson:2016fi,Sampson:2014qqa,PhysRevLett.70.2217,PhysRevD.48.3436}). Treating the energy densities in this manner, however, forces sharp transitions between different cosmological eras (i.e.~from matter domination to dark energy domination). To avoid any sharp transition and ensure a smooth evolution, we simply consider all energy components of the universe for all time and evolve the full form of Eqs.~(\ref{eq:scalar_evo})--(\ref{eq:c3}) numerically.

\begin{figure*}[t]
		\centering
		\includegraphics[width=3.5in]{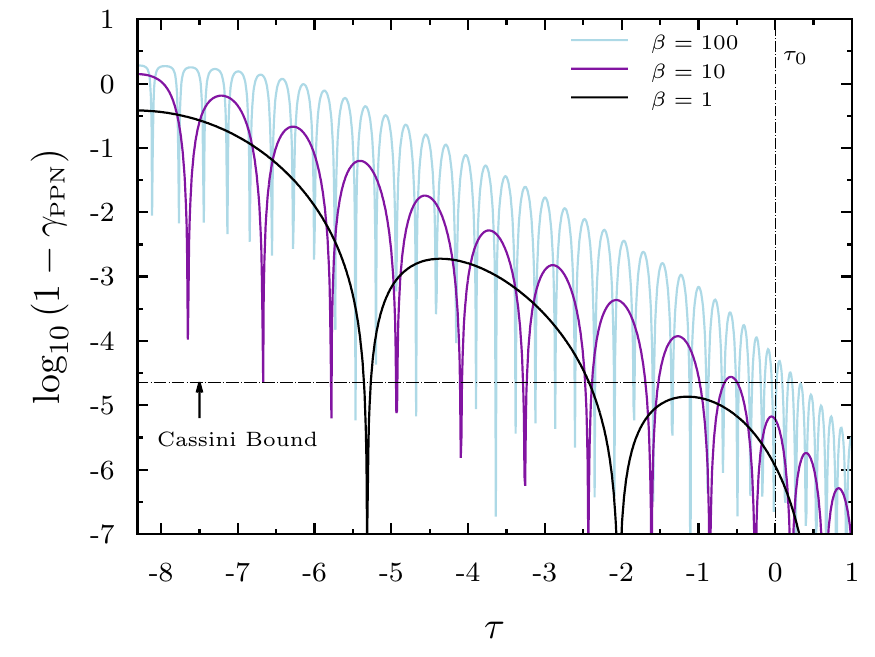}
		\includegraphics[width=3.5in]{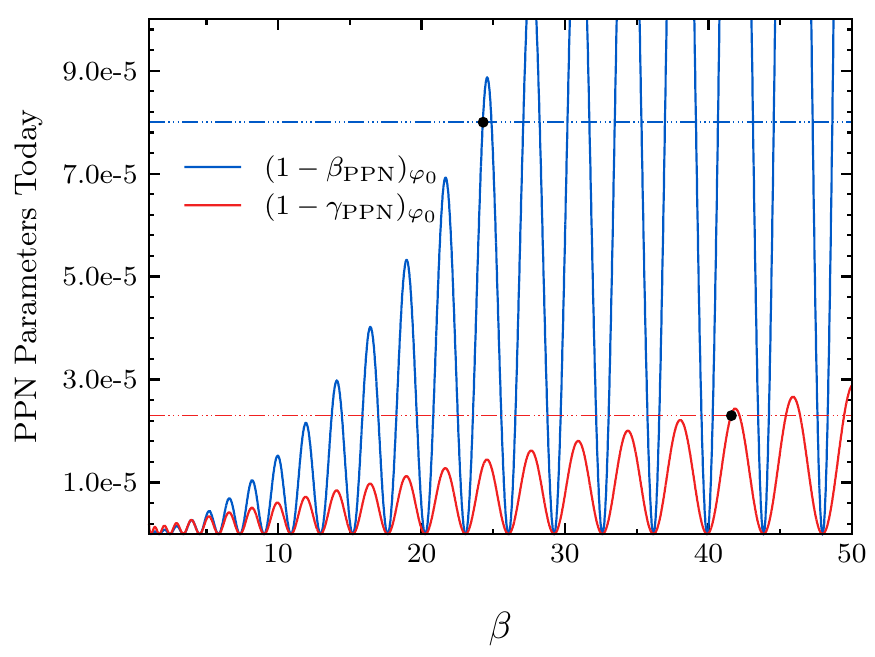}
		\caption{ \label{fig:gamma_ppn_DEF} (Color Online) \textbf{Left:} $1-\gamma_\ppN$ evaluated during the evolution of the scalar field, with $\zeta=1$ here, to show that an increase in $\beta$ makes it harder to satisfy the Cassini bound today at $\tau_{0}=0$. \textbf{Right:} $1-\gamma_\ppN$ (red) and $1-\beta_\ppN$ (blue) evaluated today at $\tau_{0} = 0$ for a range of $\beta$.
		}
	\end{figure*}

The left panel of Fig.~\ref{fig:gamma_ppn_DEF} shows the evolution of $1-\gamma_\ppN$ in the exponential theory for the entire integration time and three different values of $\beta > 0$.
	A general feature of the evolution regardless of $\beta$ is that there are fast oscillations inside of a decaying envelope, which should not come as a surprise. The scalar field starts at some height in the potential and oscillates about the minimum. Because there is a damping term in the evolution equation, the amplitude of the oscillations must decrease in time. The dips in this plot are points when the scalar field passes through the minimum of the potential, i.e.~when $\alpha(\varphi)=0$, and the peaks correspond to turning points in the evolution where $\varphi'$ instantaneously vanishes. 
	
	There are a few features of note that are present for different values of $\beta$. As $\beta$ increases, the period of oscillations about the minimum becomes smaller and therefore the motion is more rapid. This is primarily due the steepness of the corresponding potential, i.e.~when $\beta$ is large the potential is steep and narrow, allowing the velocity to be large. This accounts for two features of the evolution: 1) the period of oscillation is shorter and 2) the damping is weaker so the decay is slower in $\tau$-time. The first point above is obvious: starting from a higher point in the potential means there is more ``potential energy'', which converts to ``kinetic energy'', leading to a faster velocity and therefore faster oscillations, but the second point may seem counterintuitive. Recalling Eq.~(\ref{eq:scalar_evo}), as $\varphi'$ becomes larger the term $3-{\varphi'}^2$ becomes smaller, meaning that the damping term proportional to $\varphi'$ alone holds less significance during the evolution. This velocity-dependent mass term is what is responsible for the decreased damping at the beginning of the evolution as $\beta$ increases.
	
	A final feature of the evolution worth noting is the different decay rates that occur in the left panel of Fig.~\ref{fig:gamma_ppn_DEF} (most apparent for $\beta=100$). One would expect to see different decay rates for different intervals of $\tau$ during the evolution because damping is determined by the value of $C_2$ in Eq.~(\ref{eq:c2}). As time passes, the dominant energy density of the universe also changes, which causes $C_2$ to be dominated by that energy and results in a distinct decay rate. Before $\tau=-6$ the effects of both radiation and matter are important, until about $\tau=-1$ matter seems to completely dominate the damping, and for  $\tau \gtrsim -0.5$ the effect of dark energy becomes dominant causing the fastest decay. Although we do no show it explicitly, similar conclusions can be drawn from the behavior of $\beta_\ppN$ throughout the evolution of the scalar field.
	
	The right panel of Fig.~\ref{fig:gamma_ppn_DEF} shows the value of $|1-\gamma_\ppN|$ (red) and $|1-\beta_\ppN|$ (blue) today for a subset of $\beta$ that we considered. As one may expect from the previous results, as $\beta$ becomes larger the damping becomes weaker and as a result the field may not settle to the minimum of the potential by today. There are, however, periodic values of $\beta$ that do allow for Solar System tests to be passed even when $\beta$ is large. These values of $\beta$ lead to an evolution where the scalar field just happens to be passing by the minimum of the potential today and thus the Solar System bounds are satisfied. However, as time continues to progress the scalar field will leave the minimum of the potential in the future and may eventually fail to satisfy Solar System constraints. Such an outcome becomes unavoidable for $\beta\gtrsim 30$. 
	
	The apparent linear $\beta$ dependence of $(1-\gamma_\ppN)_{\varphi_0}$ can be explained by an approximate analytic solution to Eq.~(\ref{eq:scalar_evo}). For simplicity, consider an era that is dominated by matter and assume $\varphi' \ll \sqrt{3}$ such that the scalar-field evolution can be approximated by
	\be
		\dfrac{2}{3}\varphi'' + \varphi' = -\beta\varphi\,\,,
	\label{eq:evo_DEF_matter}
	\ee
	which is a classical damped harmonic oscillator. With the initial conditions in Eqs.~(\ref{eq:DEF_ICs1})-(\ref{eq:DEF_ICs2}), with $\zeta=1$, the solution becomes
	\be
		\varphi(\tau) = \varphi_{\mbox{\tiny $R$}} \,e^{-T(\tau)}\left[\cos(B\,T(\tau)) + \dfrac{\sin(B\,T(\tau))}{B}\right]
	\label{eq:scalar_analytic}
	\ee
	where $T(\tau) = 3(\tau+\tau_{\mbox{\tiny $R$}})/4$, $B = \sqrt{(8/3)\beta -1}$, and $\tau_{\mbox{\tiny $R$}} = -8.306$. Note that this solution is valid only for the under-damped case such that $\beta > 3/8$, which is precisely the region of parameter space we are interested in. For a more detailed discussion of the critically- and over-damped cases see Ref.~\cite{PhysRevD.48.3436}. Because $\alpha^2 = \beta^2 \varphi^2$ is small today (it is of the same order as $1-\gamma_\ppN$ in most cases) we can approximate Eq.~(\ref{eq:gamma_ppn}) as
	\be
		1-\gamma_\ppN = 2\alpha^2 = 2\beta^2\varphi^2\,\,.
	\ee
	Evaluating $\varphi$ in Eq.~(\ref{eq:scalar_analytic}) today and averaging over the small oscillations to extract the secular terms, one finds 	
	\be
		(1-\gamma_\ppN)_{\varphi_0} \,\propto\, \beta^2\varphi_{\mbox{\tiny $R$}}^2 \,\propto\, \beta\,\,,
	\ee
	using Eq.~(\ref{eq:DEF_ICs1}) for $\varphi_{\mbox{\tiny $R$}}$. In other words, the overall growth of $1 - \gamma_\ppN$ today is linear in $\beta$, as we see in the right panel of Fig.~\ref{fig:gamma_ppn_DEF}. Likewise, $1 - \beta_\ppN$ has an extra factor of $\beta$ in the numerator, making it quadratic, and thus grows even faster than $1 - \gamma_\ppN$, leading to a stronger violation of Solar System constraints for large $\beta$.

\begin{figure*}[t]
	\centering
	\includegraphics[width=3.5in]{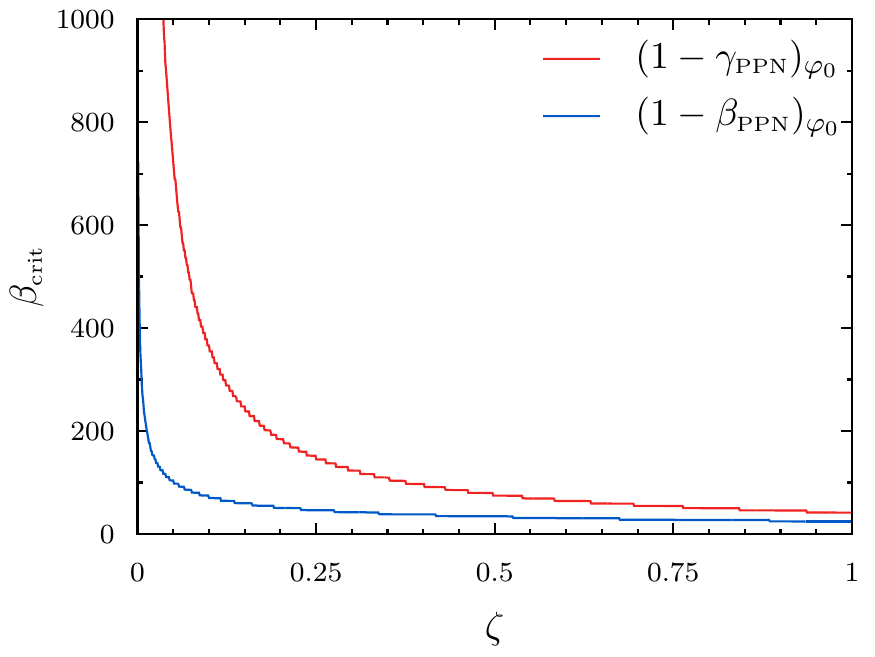}
	\includegraphics[width=3.5in]{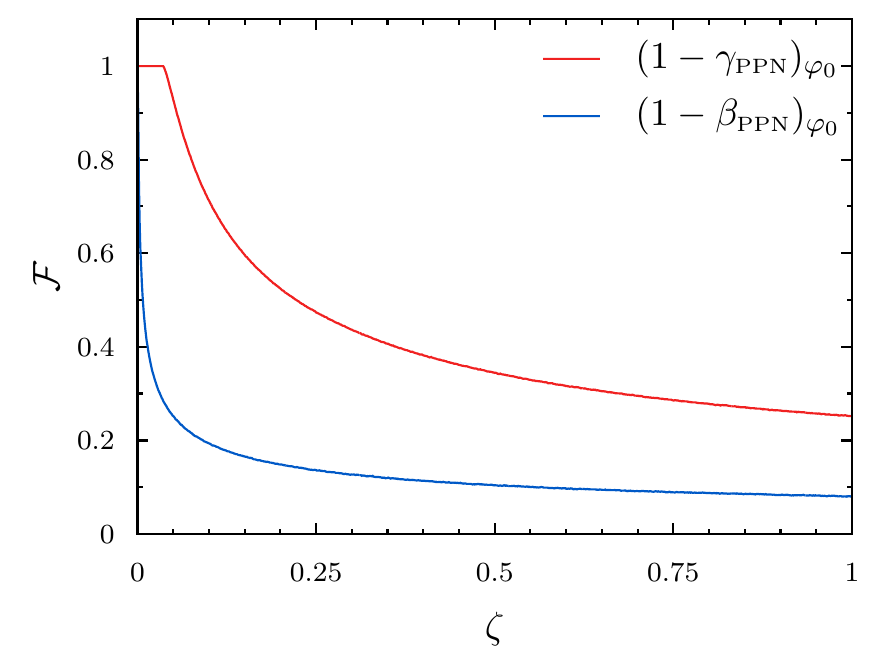}
	\caption{ \label{fig:crit_frac_DEF} (Color Online) \textbf{Left:} Value of $\beta_\crit$, as defined in the text, for every value of $\zeta$ which corresponds to the the entire range of possible initial scalar positions at the time of BBN. \textbf{Right:} The total fraction out of all $\beta$ values sampled ($1<\beta<1000$) that allow Solar System tests to be passed at $\tau=0$, as a function of the initial conditions determined by $\zeta$. The general trend of the top right panel tells us that this fraction would continue to decrease as we increase the upper bound of our sample size.
	}
\end{figure*}
\subsubsection{Constraints on $\beta$}\label{sec:beta_constriants}
	
	We now turn our attention to placing quantitative constraints on $\beta$ from the results of our numerical calculations. Because of the oscillatory nature of the scalar field, there are different types of constraints that can be placed, which differ in their generality: 
	\begin{itemize}
		\item $\beta_\crit$, where all $\beta<\beta_\crit$ pass Solar System at the present time but not necessarily at all future times,
		\item $\beta_\fail$, where all $\beta<\beta_\fail$ pass Solar System tests at the present time, and generically also at all future times (this is the method used in Ref.~\cite{Anderson:2016fi}),
		\item $\beta_\ffail$, where \emph{all} $\beta>\beta_\ffail$ will generically fail Solar System tests either today or at some future time.
	\end{itemize}
	All of these possibilities lead to constraints that are consistent with each other, but for completeness we present all of them in our analysis. Furthermore, we will show below that these constraints are relatively insensitive to the value of $\zeta$, i.e.~the initial conditions chosen at the time of BBN, provided that they are not finely tuned to be close to $\zeta=0$ $(\varphi=0)$.
	
	The discussion in the previous section specifically applied to the $\zeta=1$ case only, so now we will focus on the consequences of allowing $0 < \zeta < 1$. For every set of initial conditions determined by $\zeta$ in Eqs.~(\ref{eq:DEF_ICs1})-(\ref{eq:COSH_ICs2}) there exists a critical value of $\beta$, denoted by $\beta_\crit$, for which all $\beta<\beta_\crit$ lead to a value of $\varphi_0$ that is consistent with Solar System observations. In general, there is a $\beta_\crit$ associated with each PPN parameter, corresponding to the intersections of the constraints (red/blue horizontal lines) and corresponding curves (red/blue curves) in the right panel of Fig.~\ref{fig:gamma_ppn_DEF}. The smaller of these two will be the value of $\beta_\crit$ reported in this paper, since it places the tightest constraint. For $\zeta=1$, one can see from the right panel of Fig.~\ref{fig:gamma_ppn_DEF} that $\beta_\crit=24$, corresponding to the black point where the blue horizontal line intersects the curve for $|1-\beta_\ppN|$.  The left panel of Fig.~\ref{fig:crit_frac_DEF} shows the behavior\footnote{The step-like nature of $\beta_{\crit}$ is related to the oscillatory behavior of the scalar field shown in Fig.~\ref{fig:gamma_ppn_DEF}, and it is not an artifact of numerical resolution.} of $\beta_\crit$ for all values of $\zeta$, associated with both $\gamma_\ppN$ and $\beta_\ppN$. Small values of $\zeta$ lead to exponentially large values of $\beta_\crit$, such that $\beta_\crit \rightarrow \infty$ in the limit that $\zeta \rightarrow 0$ for both PPN parameters. This should indeed be the case because when $\zeta \rightarrow 0$ the scalar field is sitting at the minimum of the potential at the time of BBN. If the field is at the minimum in the past, with no velocity, then it will certainly be at the minimum today, and thus it will satisfy Solar System tests with ease. Our results show that regardless of the value of $\zeta$, current constraints on $\beta_\ppN$ place the tightest constraints on the value of $\beta_\crit$. The values of $\beta_\fail$ and $\beta_\ffail$ are determined in a similar fashion and have a very similar dependence on $\zeta$ as $\beta_\crit$ does. We list all  of these values in Table~\ref{tab:betas}.
	
	Ultimately we want to place bounds on values of $\beta$ and use those in our NS calculations, but considering $\zeta\sim 0$ does not restrict $\beta$ at all according to the arguments presented above. We therefore restrict our calculations to $\zeta>0.05$ as a way of avoiding a large amount of fine-tuning; the choice of 0.05 as the lower bound is because in a randomly selected sample of $\zeta$ there is only a 5\% chance that $\zeta<0.05$ would be selected. In another effort to avoid a finely tuned bound, we take a random distribution of $\zeta$ and select the median of the resulting distribution of $\beta_\crit$ to represent a measure of the ``most probable'' value of $\beta_\crit$, denoted by $\beta_{\mbox{\tiny crit,med}}$,  if $\zeta$ were chosen randomly. We find that in the exponential theory this median value is $\beta_{\mbox{\tiny crit,med}}\sim 34$. 
	
	The quantities $\beta_\crit$ and $\beta_{\mbox{\tiny crit,med}}$ still do not necessarily tell the entire story of which values of $\beta$ should be considered feasible because, as Fig.~\ref{fig:gamma_ppn_DEF} shows, there is a (small) subset of $\beta>\beta_\crit$ (or $\beta>\beta_{\mbox{\tiny crit,med}}$ in the case of more finely-tuned initial conditions) that do indeed pass Solar System tests today. To quantify this, we define
	\be
		\mathcal{F} \eq \dfrac{\# \text{ of }\beta \text{ that pass Solar System tests}}{\# \text{ of }\beta \text{ sampled}}
	\ee
	and present these results as a function of $\zeta$ in the  right panel of Fig.~\ref{fig:crit_frac_DEF}. As a reminder, $\zeta=1$ corresponds to the solutions displayed in Fig.~\ref{fig:gamma_ppn_DEF} where $\beta_\crit \sim 24$, as determined by constraints on $\beta_\ppN$, and in this case $\mathcal{F}$ is only slightly over 8\%. This means that there is roughly an 8\% chance that a randomly selected value of $\beta$ between 1 and 1000 will allow Solar System tests to be passed. Interestingly, $\mathcal{F}$ changes very slowly with $\zeta$, such that even for very small values of $\zeta$, i.e.~for $\zeta > 0.05$, $\mathcal{F}$ is still $\lesssim 25\%$.
		
	The above results may seem to indicate that there is a decent number of possible values of $\beta$ above the critical or the median one for which Solar System tests are passed, provided that we sample larger values of $\beta$, but this conclusion would be premature. We have here greatly narrowed the scope of our study by only focusing on $\beta < 1000$, but in principle, $\beta$ could take on any value between zero and infinity. Considering values of $\beta>1000$, say up to $\beta=10^6$ for definiteness, would lead to two consequences: (i) the \# of $\beta$ that pass Solar System tests would increase slightly, and (ii) the \# of $\beta$ sampled would increase dramatically. Because of these effects, the relative fraction $\mathcal{F}$ would decreases significantly leaving a very small probability of passing Solar System tests without a very fine-tuned choice of $\beta$.

	\begin{figure*}[t]
		\centering
		\includegraphics[width=3.5in]{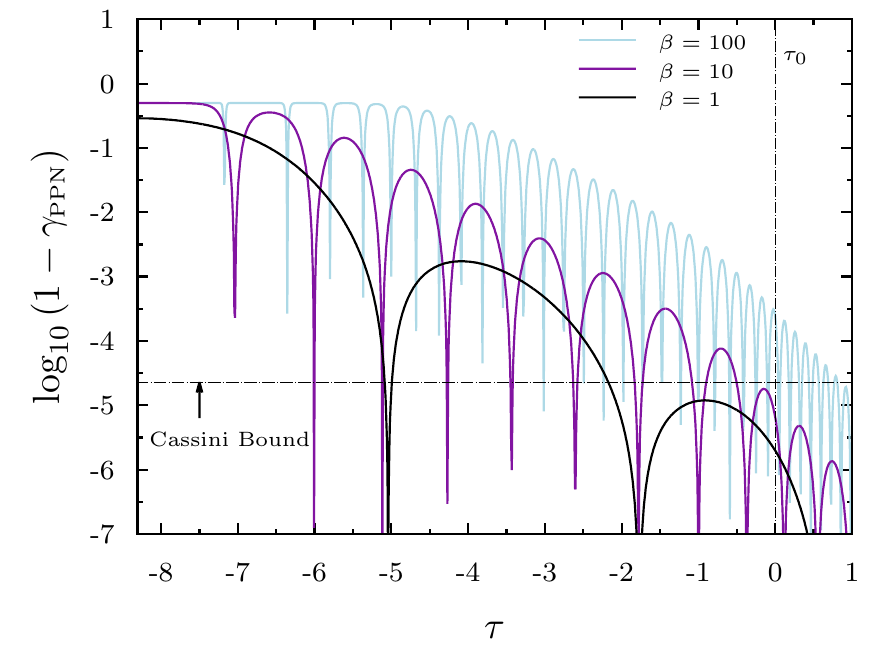}
		\includegraphics[width=3.5in]{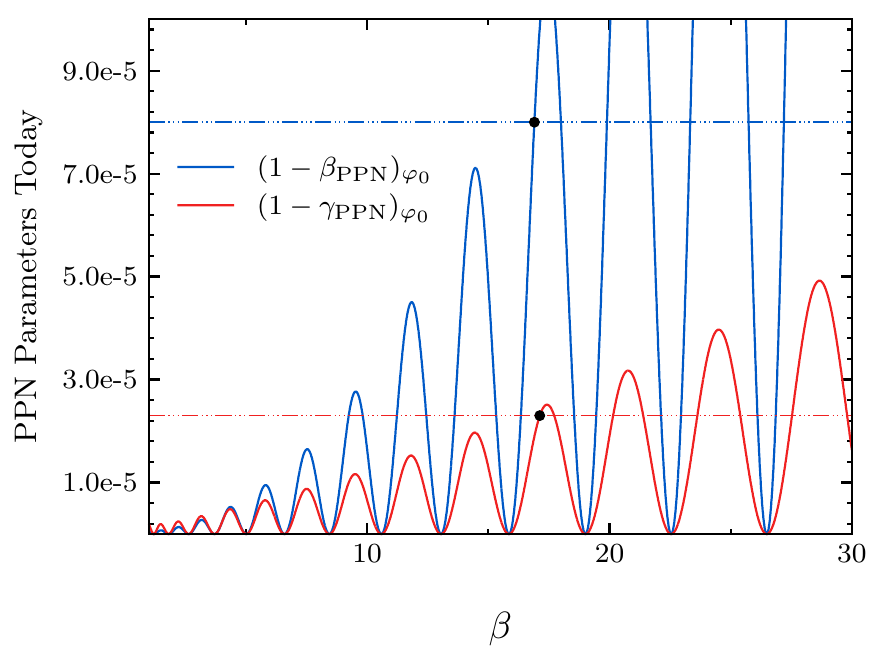}
		\includegraphics[width=3.5in]{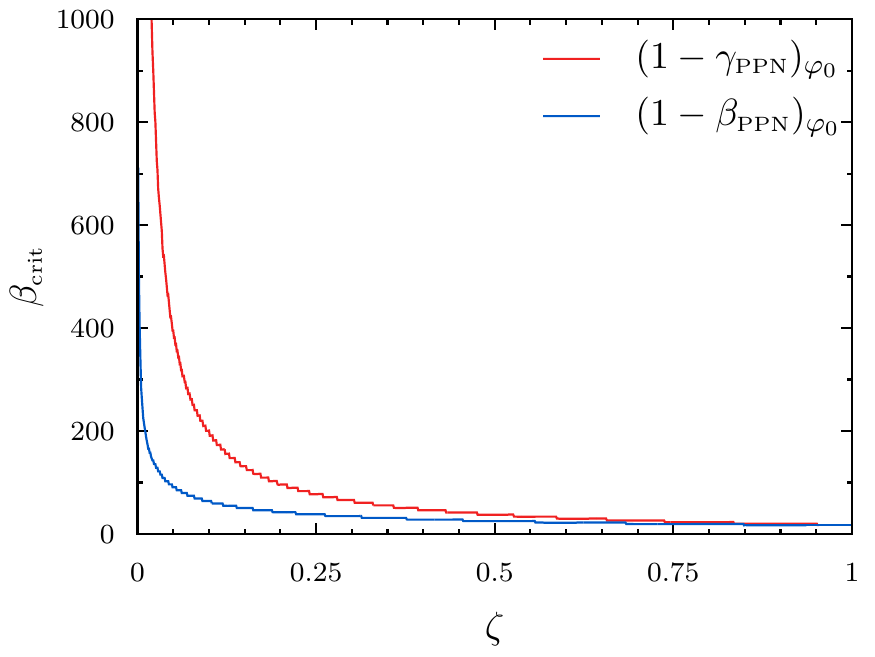}
		\includegraphics[width=3.5in]{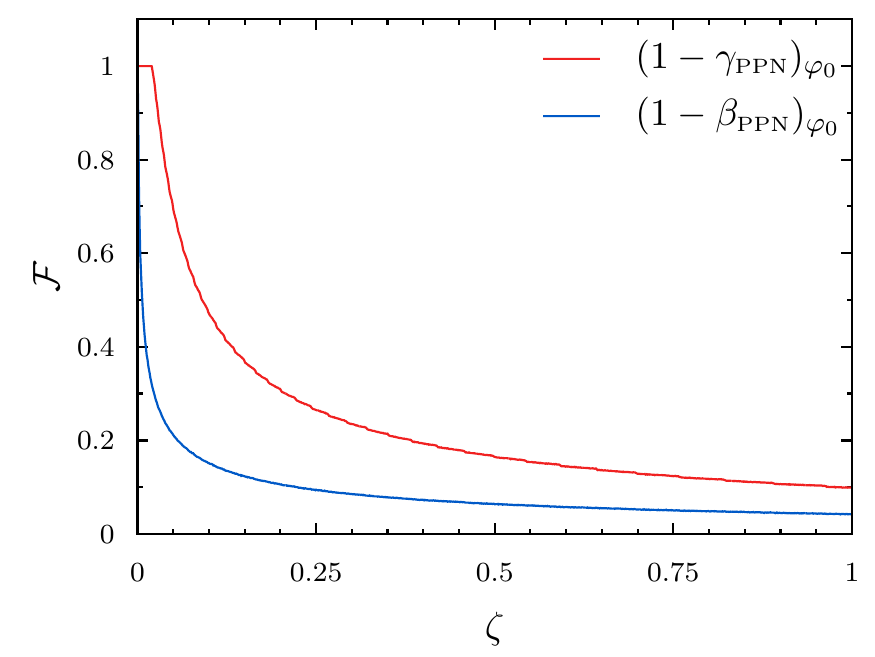}
		\caption{ \label{fig:gamma_ppn_COSH} (Color Online) Same as Fig.~\ref{fig:gamma_ppn_DEF} and~\ref{fig:crit_frac_DEF} but for the hyperbolic class of STTs.
		}
	\end{figure*}
	%
	%
	
	Let us wrap up this subsection by summarizing the main results we have obtained, all of which can be inferred from Fig.~\ref{fig:gamma_ppn_DEF} and Fig.~\ref{fig:crit_frac_DEF}. For the $\zeta=1$ case, we see that for all $\beta \lesssim \beta_\crit \sim 24$ the scalar field will evolve such that Solar System tests are passed today. For $\beta \gtrsim \beta_\crit \sim  24$ the solutions for which Solar System tests are passed is periodic in nature, but in general they do not pass Solar System tests. Moreover, the ones that do pass Solar System tests today do no not generically do so in the future. In fact, all $\beta \gtrsim 29$ that happen to pass Solar System tests today will eventually fail them in the future.  The monotonic growths seen in the right panel of Fig.~\ref{fig:gamma_ppn_DEF} imply that this trend will continue as $\beta$ increases and there will never be a turnover such that larger ranges of $\beta$ will generally pass Solar System tests. Such conclusions turn out to be valid for all values of $\zeta>0$. A summary of these results can be found in Table~\ref{tab:betas}.

\subsection{Hyperbolic Theory}
	We now repeat the analysis described above but using the hyperbolic theory instead of the exponential one. The numerical results for the evolution of the scalar field are displayed in Fig.~\ref{fig:gamma_ppn_COSH}. There are several similar features in the scalar field evolution to what we found in the exponential case, as well as a few features exclusive to hyperbolic coupling that we discuss below.

	The upper panels of Fig.~\ref{fig:gamma_ppn_COSH} show the evolution of the scalar field for $\zeta =1$. As $\beta$ increases, there is a distinct increase in the frequency of oscillation about the minimum of the potential. In the exponential theory, this was attributed to the fact that the potential became steeper with increasing $\beta$, generating steeper gradients and therefore larger accelerations. The potential gradient $\alpha(\varphi)$ in the hyperbolic theory exponentially asymptotes to a value of $1/\sqrt{3}$ for large values of $\beta\varphi$, cf. Fig.~\ref{fig:potentials_couplings}. This behavior limits the maximum value of $1-\gamma_\ppN$ via Eq.~(\ref{eq:gamma_ppn}), and also restricts the coupling between the scalar field and matter, effectively placing an upper limit on the sourcing term of the oscillator. As a result, the scalar field does not gain as much velocity as quickly here, and therefore, it spends more time away from the minimum of the potential initially, leading to minimal amounts of oscillatory motion and the saturation of $1-\gamma_\ppN$ occurring in Fig.~\ref{fig:gamma_ppn_COSH} at early times. Again, while we do not show the time evolution of $1-\beta_\ppN$, its behavior is similar and ultimately leads to identical qualitative conclusions.

	Once enough time has passed and the scalar field has been damped enough to remain near the minimum of the potential, it finally begins to behave as in the exponential theory, but this has a significant effect on whether or not Solar System tests are passed today. Because $\alpha(\varphi)$ remains at its limiting value for a substantial amount of time, which is evident for example when $\beta>10$, the friction terms do not have a significant effect until later in the evolution, making it harder for the field to decay to a small enough value in order for either PPN parameter to be consistent with Solar System constraints. This is made clear in the top, right panel of Fig.~\ref{fig:gamma_ppn_COSH} where we see that $\beta_\crit \sim17$, a value smaller than that found in the exponential case. We also see that it becomes even more difficult to pass Solar System tests for large $\beta$, i.e. the regions falling below the Solar System bounds become smaller the larger $\beta$ becomes. We also find that all $\beta \gtrsim 19$ that allow Solar System tests to be passed today will ultimately fail to do so at some point in the future.

	Similar to the analysis in the previous section we can study what effect the initial conditions, parameterized via $\zeta$, have on the scalar-field evolution, with the bottom panels of Fig.~\ref{fig:gamma_ppn_COSH} showing $\beta_\crit$ and $\mathcal{F}$ as functions of $\zeta$. The overall behavior of $\beta_\crit$ is similar to that in the exponential theory but it is always significantly lower. Just like we did previously, we can quantify a median value of $\beta_\crit$ corresponding to the median of the distribution appearing in the bottom, left panel of Fig.~\ref{fig:gamma_ppn_COSH} and we find that $\beta_{\mbox{\tiny crit,med}} \sim 25$ for the hyperbolic theory. The fraction $\mathcal{F}$ is also smaller and is only around half of what it is in the exponential theory. As we argued before, this fraction will only decrease with an increased sample size of $\beta$, and to avoid fine-tuning scenarios one should try to avoid considering value of $\beta$ larger than $\beta_\crit$ that satisfy Solar System tests today because they just happen to have $\alpha(\varphi)=0$ today. Table~\ref{tab:betas} contains a summary of all bounds on $\beta$ from these calculations.

\subsection{Constraints on $\beta$: Summary}

	\begin{table*}[t]
		\centering 
		\renewcommand{\arraystretch}{1.3}
		\begin{tabular}{!{\vrule width 1pt}C{0.15\linewidth}| L{0.15\linewidth}!{\vrule width 1pt} C{0.15\linewidth}| C{0.15\linewidth}!{\vrule width 1pt} C{0.15\linewidth}| C{0.15\linewidth}!{\vrule width 1pt} C{0.01\linewidth}}
			\Cline{1pt}{3-6}
			\multicolumn{2}{C{0.3\linewidth}!{\vrule width 1pt}}{}	&	\multicolumn{2}{C{0.3\linewidth}!{\vrule width 1pt}}{Exponential}	&	\multicolumn{2}{C{0.3\linewidth}!{\vrule width 1pt}}{Hyperbolic}	&\\ \cline{3-6} 
			\multicolumn{2}{C{0.3\linewidth}!{\vrule width 1pt}}{}	&	$(1-\gamma_\ppN)_{\varphi_0}$&	$(1-\beta_\ppN)_{\varphi_0}$	&	$(1-\gamma_\ppN)_{\varphi_0}$&	$(1-\beta_\ppN)_{\varphi_0}$	&\\ \Cline{1pt}{1-6}
			\multirow{4}{2cm}[.1 cm]{\centering\bf{$\zeta = 0.05$}}	&	\hspace{0.4\linewidth}$\beta_\crit$	&	722	&	104	&	395	&	91	& \\[-2pt] 
			&	\hspace{0.4\linewidth}$\beta_\fail$	&	722	&	103	&	394	&	91& \\[-2pt] 
			&	\hspace{0.4\linewidth}$\beta_\ffail$	&	811	&	113	&	431	&	100	& \\[-2pt]  
			&	\hspace{0.4\linewidth}$\mathcal{F}$	& 93\%	&	25\%	&	69\%	&	21\%	& \\[-2pt]  \Cline{1pt}{1-6}
				\multirow{4}{2cm}[.1 cm]{\centering\bf{$\zeta = 1$}}	&	\hspace{0.4\linewidth}$\beta_\crit$	&	42	&	24	&	17	&	17	& \\[-2pt]
			&	\hspace{0.4\linewidth}$\beta_\fail$	&	41	&	24	&	17	&	17& \\[-2pt] 
			&	\hspace{0.4\linewidth}$\beta_\ffail$	&	60	&	29	&	25	&	19	& \\[-2pt] 
			&	\hspace{0.4\linewidth}$\mathcal{F}$	& 25\%	&	8\%	&	10\%	&	4\%	& \\[-2pt]  \Cline{1pt}{1-6}
			\multirow{2}{2cm}[0 cm]{\centering{Median Values}}	&	\hspace{0.4\linewidth}$\beta_\crit$	&	76	&	34	&	38	&	25	& \\[-2pt] 
			&	\hspace{0.4\linewidth}$\beta_\ffail$	&	99	&	40	&	49	&	27& \\  \Cline{1pt}{1-6}
			\multicolumn{2}{!{\vrule width 1pt}C{0.3\linewidth}!{\vrule width 1pt}}{\multirow{2}{2cm}[0 cm]{$\beta_\crit=50$}}	&	$\zeta<0.843$	&	$\zeta<0.229$	&	$\zeta<0.393$	&	$\zeta<0.162$	& \\ 
			\multicolumn{2}{!{\vrule width 1pt}C{0.3\linewidth}!{\vrule width 1pt}}{}&	$\varphi_{\mbox{\tiny $R$}}<0.063$	&	$\varphi_{\mbox{\tiny $R$}}<0.032$	&	$\varphi_{\mbox{\tiny $R$}}<0.088$	&	$\varphi_{\mbox{\tiny $R$}}<0.041$	& \\\cline{1-6}
			\multicolumn{2}{!{\vrule width 1pt}C{0.3\linewidth}!{\vrule width 1pt}}{\multirow{2}{2cm}[0 cm]{$\beta_\crit=100$}}	&	 $\zeta<0.376$	&	$\zeta<0.051$	&	$\zeta<0.196$	&	$\zeta<0.043$& \\
			\multicolumn{2}{!{\vrule width 1pt}C{0.3\linewidth}!{\vrule width 1pt}}{}&	 $\varphi_{\mbox{\tiny $R$}}<0.029$	&	$\varphi_{\mbox{\tiny $R$}}<0.011$	&	$\varphi_{\mbox{\tiny $R$}}<0.044$	&	$\varphi_{\mbox{\tiny $R$}}<0.013$& \\ \Cline{1pt}{1-6}
		\end{tabular}
		\caption{\label{tab:betas} Bounds on $\beta$ using different initial conditions (determined by $\zeta$) and both PPN parameters ($\gamma_\ppN$ and $\beta_\ppN$) in both the exponential and hyperbolic theories. The parameters $\beta_\crit$, $\beta_\fail$, $\beta_\ffail$, and $\mathcal{F}$ are those defined in the text. The first row shows data for lowest value of $\zeta$ we consider, i.e.~$\zeta=0.05$. The second row show the data for saturating BBN constraints ($\zeta=1$). We present the median values of certain parameters in the third row, given by the definitions defined in the text. The last two rows illustrate the amount of fine tuning of $\zeta$, or alternatively $\varphi_{\mbox{\tiny {$R$}}}$, that would be necessary to force $\beta_\crit$ to either 50 or 100.  
		}
	\end{table*}
	Let us now briefly summarize our conclusions regaring the bounds we can place on $\beta$ given Solar System observations and their implications. First, recall that we focused on the range $1<\beta<1000$ and $0<\zeta<1$, and that $\zeta$ close enough to zero, e.g.~$\zeta <0.05$, corresponds to a finely tuned choice of initial conditions, i.e.~choosing $\varphi\sim 0$ at the time of BBN. Table~\ref{tab:betas} shows representative values of the bounds on $\beta$ from the three choices discussed at the beginning of Sec.~\ref{sec:beta_constriants}, for both PPN parameters in both the exponential and hyperbolic theories. Notice that in every case displayed, the bounds placed from $\beta_\ppN$ are all tighter than the ones placed by $\gamma_\ppN$. This is consistent with the structure of the PPN constraints because $\beta_\ppN$ takes the curvature of the conformal potential into account, while $\gamma_\ppN$ does not. For any $\beta \gtrsim 10$ the curvature of the potential becomes significant near the minimum of the potential, in turn causing larger present day values of $|1-\beta_\ppN|$ which are inconsistent with Solar System observations. We therefore focus explicitly on the bounds placed by $\beta_\ppN$.
	
	Let us first focus on the $\zeta=1$ portion of Table~\ref{tab:betas}. For both theories, $\beta_\crit$, $\beta_\fail$, and $\beta_\ffail$ all place nearly identical upper bounds on $\beta$. Furthermore, $\mathcal{F}$ is $8\%$ and $4\%$ for the exponential and hyperbolic theory respectively. Such small values of $\mathcal{F}$ suggest that it is extremely hard to randomly select a value of $\beta>\beta_\crit$ that will be consistent with Solar System constraints, and in fact this value only becomes smaller by considering values of $\beta>1000$. This suggest that one would have to very precisely select a $\beta>\beta_\crit$ if one wants to remain consistent with Solar System observations, leading to a fine-tuning problem. Therefore, in order to naturally avoid this, one should only consider $\beta<\{\beta_\crit,\,\beta_\fail,\,\beta_\ffail\}$, all of which are $\sim 20$ in both theories.
	
	Let us now discuss the fine-tuning issue in a bit more detail, by focusing on the rest of Table~\ref{tab:betas}. First, notice that considering $\zeta=0.05$ brings the upper bounds on $\beta$ up to $\sim 100$ and $\mathcal{F}$ up to $\sim 25\%$ for both theories. While this increases the viable range of $\beta$, one should remember that forcing $\zeta$ to be small is equivalent to forcing $\varphi \rightarrow 0$ at the time of BBN, resulting in a carefully selected set of initial conditions. To avoid this fine tuning issue, we calculated the median value of the distribution of $\beta_\crit$ and $\beta_\ffail$, which represent the $\beta$'s that would be \emph{likely} to pass Solar System tests if $\zeta$ were chosen randomly (these are located in the rows label ``Median Values'' in Table.~\ref{tab:betas}). Ultimately, we see that this type of analysis only slightly relaxes the bounds on $\beta$ relative to what is found when $\zeta=1$, meaning that for a large subset of $0<\zeta<1$ the bounds on $\beta$ are around $\sim 40$ and $\sim 25$ for the exponential and hyperbolic theory respectively.
	
	Lastly, let us quantify the amount of fine-tuning that would be needed in order to raise $\beta_\crit$ to a region containing non-zero scalar charge in Fig.~\ref{fig:max_charge_beta}. This information is shown in the bottom two rows of Table~\ref{tab:betas}, but let us here only focus on the hyperbolic theory since we only find stable NSs with non-zero scalar charge there. In order to raise $\beta_\crit$ to 50 (100), $\zeta$ would have to be smaller than 0.162 (0.043). To achieve this one needs to demand that the scalar field be close to the minimum of the conformal potential and therefore be very similar to GR on cosmological scales. While allowing STTs to reduce to GR ($\zeta \sim 0$) on these scales would allow Solar System tests to be passed it is effectively a set of measure zero. For the reasons above, we conclude that the most natural way to evade Solar System constraints today is to consider bounds on $\beta$ corresponding to $\zeta=1$, and we note that only under a carefully selected set on initial conditions at BBN and $\beta$ is it possible to pass Solar System tests otherwise.
\section{Neutron Stars and Scalarization}
\label{Neutron Stars and Scalarization}

	In this section we describe NSs in STTs and introduce the basics of scalarization. To reiterate the argument made in  Ref.~\cite{Anderson:2016fi}, we only study isolated NSs here because theories that do not allow for spontaneous scalarization are also not likely to allow for dynamical/induced scalarization. We first present the equations of structure and then describe the piecewise polytropic EoSs we use to model realistic tabulated EoSs~\cite{Read:2009jt}. Finally, once the framework for the numerics has been laid out we present results for NSs in the exponential theory and the hyperbolic theory, first for $\beta<0$ and then for $\beta>0$.
	
	For our study, we focus explicitly on isolated, static, and spherically symmetric spacetimes, which in general can be described by the Einstein-frame line element
	\be
	ds^2_* = -e^{\nu(r_*)}dt^2_* + \frac{dr^2_*}{1-2\mu(r_*)/r_*} + r^2_*d\Omega^2_*\,\,,
	\label{eq:matric_ansatz}
	\ee
	where $\nu$ and $\mu$ are only functions of the Einstein-frame coordinate radius $r_*$ and $d\Omega^2_*$ is the line element on the two-sphere. To a good approximation, the matter inside cold NSs can be represented as a perfect fluid and therefore we use the form of the matter stress-energy tensor given in Eq.~(\ref{eq:def-perfect_fluid}). Applying conservation of stress energy in the Jordan frame, $\nabla_\nu T^{\mu\nu} =0$, along with the line element and perfect fluid assumptions above yield a set of first order differential equations governing the structure of the NSs
	\ba
	\mu' &=& 4\pi G r_*^2 A^4(\varphi)\epsilon + \dfrac{1}{2}r_*(r_*-2\mu)\psi^2\,\,,
	\label{eq:NSstructure_mu}
	\\
	\nu' &=& r_*\psi^2 + \dfrac{1}{r_*(r_*-2\mu)}\big[2\mu + 8\pi G r_*^3 A^4(\varphi) p\big]\,\,,
	\label{eq:NSstructure_nu}
	\\
	\varphi' &=& \psi\,\,,
	\label{eq:NSstructure_phi}
	\\
	\psi' &=& \dfrac{4\pi G r A^4(\varphi)}{(r_*-2\mu)}\big[ \alpha(\varphi) (\epsilon - 3p) + r_*\psi(\epsilon - p)\big]\label{eq:NSstructure_psi}\\
	&\,&\,\,\,-2\psi\dfrac{(1-\mu/r_*)}{(r_*-2\mu)}	\,\,,\nn
	\\
	p' &=& -(\epsilon + p)\left( \nu'/2 + \alpha(\varphi)\psi\right)\,\,,
	\label{eq:NSstructure_p}
	\ea
	where primes denote derivatives with respect to $r_*$, and all occurrences of $p$ and $\epsilon$ are explicitly in the Jordan frame. \footnote{There is a missing factor of $r_*$ in the Eq.~(\ref{eq:NSstructure_psi}) equivalent appearing in Reference~\cite{Anderson:2016fi}.}
	\begin{table*}[t]
		\centering
		\begin{tabular*}{7in}{L{.535 in} C{.75 in} C{.75 in} C{.75 in} C{.75 in} C{.75 in} C{.75 in} C{.75 in} C{.75 in}  L{1 pt}}
			\hline
			\hline
			EoS& $\log (p_1)$ & $\Gamma_1$ & $\Gamma_2$ & $\Gamma_3$ & $\rho_\crit/\rho_0$ & $\beta_{\mbox{\tiny min}}$ & $\beta_{\mbox{\tiny max}}$ &$\alpha_{\scm}$& \\ [5 pt]
			\hline
			ENG& 34.437 & 3.514 & 3.130 & 3.168 & 4.544  & 19.1 & 26.7 & $\sim 1.4 \times 10^{-3}$ &\\ [2 pt]
			MPA1& 34.495 & 3.446 & 3.572 & 2.887& 3.683 & 19.3 & 27.2 & $\sim 1.4 \times 10^{-3}$ &\\ [2 pt]
			MS1& 34.858 & 3.224 & 3.033 & 1.325& 3.067 & 35.6 & 46.3 & $\sim 2.7 \times 10^{-4}$ &\\ [2 pt]
			MS1b& 34.855 & 3.456 & 3.011 & 1.425& 3.092 & 35.8 & 48.3 & $\sim 2.4 \times 10^{-4}$ &\\ [2 pt]
			SLy& 34.384 & 3.005 & 2.988 & 2.851& 5.349 & 30.0 & 39.3 & $\sim 4.2 \times 10^{-4}$ &\\ [2 pt]
			\hline
		\end{tabular*}
		\caption{\label{tab:eos} Parameters of the piecewise polytropes \{$\log(p_1),\,\Gamma_1,\,\Gamma_2,\,\Gamma_3$\} for the 5 EoSs we use for NS calculations. For clarity, all $\rho>\rho_\crit$ allow $T_\mat>0$ in the core, $\beta_{\min}$ is the lowest value of $\beta$ allowing for stable scalarized solutions, $\beta_{\max}$ is where the maximal stable scalar charge occurs, and $\alpha_{\scm}$ is the maximal scalar charge, corresponding to the peak values of the curves in Fig.~\ref{fig:max_charge_beta}.
			}
	\end{table*}
	
	To close the system of equations we need to prescribe an EoS. We use the piecewise polytropic EoSs studied in Ref.~\cite{Read:2009jt} in which the authors developed a four parameter polytopic model that accurately approximates 34 tabulated EoSs. Let us briefly discuss this parameterization of the EoS. In each baryonic density region inside the star $\rho_{i-1} < \rho < \rho_i$ the pressure is given by the standard polytropic form
	\be
		p \eq K_i \rho^{\Gamma_i}\,\,,
	\ee
	where $\Gamma_i$ is the adiabatic index of the fluid in that region and $K_i$ is the polytropic constant chosen to ensure continuity at the boundary between density regions. The energy density $\epsilon$ appearing in the structure equations is then found by integrating the first law of thermodynamics
	\be
		d\dfrac{\epsilon}{\rho} \eq -p\,d\dfrac{1}{\rho}\,\,,
	\label{eq:thermo1}
	\ee
	to find
	\be
		\epsilon \eq \rho + \dfrac{p}{\Gamma - 1}\,\,.
	\label{eq:eps-rho}
	\ee
	%
	%
	
	We adopt a low-density crust region described by a degenerate relativistic gas with adiabatic constant $\Gamma_0 = 4/3$ and match it to a polytrope with adiabatic index $\Gamma_1$. The polytropic constant $K_0$ for this crust region is determined by forcing the crust to match the SLy EoS at a baryonic density of $\rho_0 = 2.7\times 10^{14}$ g/cm$^3$. A different choice of a fixed crust EoS only affects the physical quantities of the system by a few percent. More importantly, scalarization will only occur in the dense regions of the star, which are unaffected by the choice of crust used, so the simple choice we make here will suffice for our study. The boundary between the crust and the next polytropic region will therefore depend on the value of $\Gamma_1$. At a fixed baryonic density of $\rho_1 = 10^{14.7}$ g/cm$^3$ and pressure $p_1=p(\rho_1)$ the EoS is matched to another polytrope with adiabatic index $\Gamma_2$ and constant $K_2$. Another boundary is set at $\rho_2 = 10^{15}$ g/cm$^3$ where the EoS is matched yet again to another polytrope with index $\Gamma_3$ and constant $K_3$. The set of parameters \{${\log(p_1),\,\Gamma_1,\,\Gamma_2,\,\Gamma_3}$\} determines the best fit to tabulated EoSs and were determined in Ref.~\cite{Read:2009jt}, which we refer the reader to for  a more detailed description. Following Ref.~\cite{Mendes:2015gx} we restrict the scope of our investigation to EoSs that allow for a maximum gravitational mass of more than two solar masses and do not violate causality. These constraints leave us with 5 EoSs, whose parameters are listed in Table~\ref{tab:eos} for convenience.

	\begin{figure*}[t]
		\centering
		\includegraphics[width=3.5in]{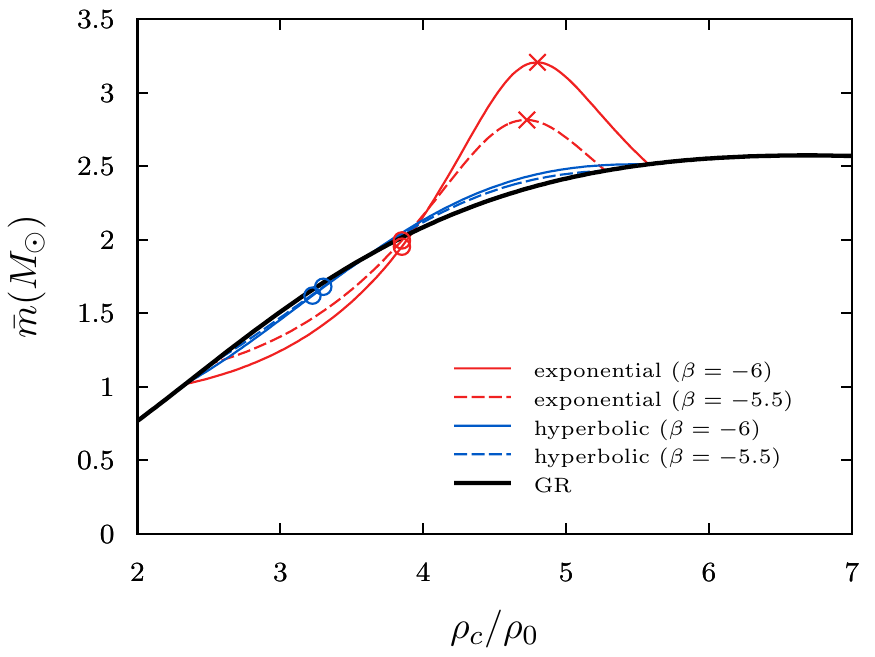}
		\includegraphics[width=3.5in]{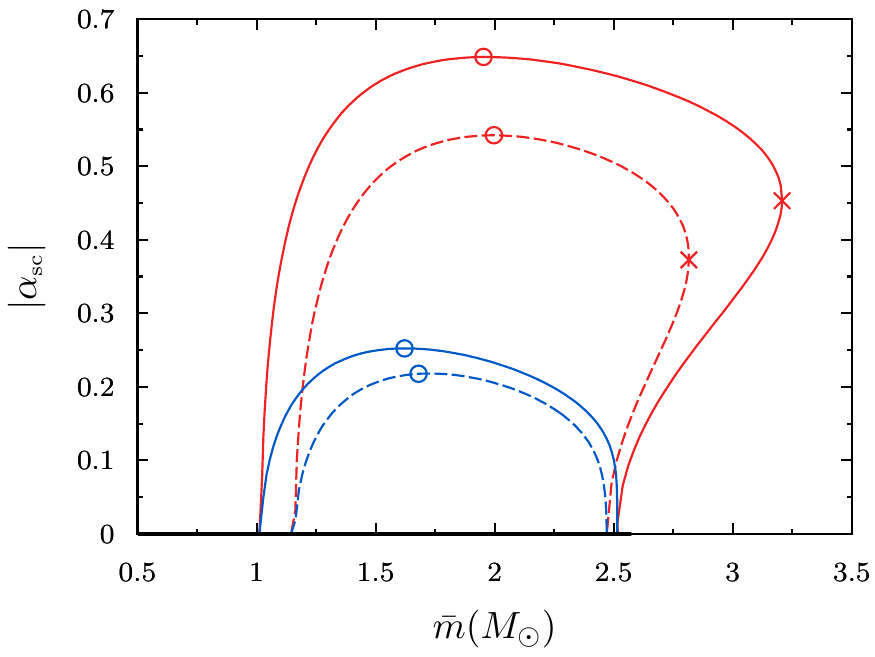}
		\caption{ \label{fig:negative_beta} (Color Online) \textbf{Left:} Baryonic mass of the NS as a function of central baryonic density. Deviations from GR occur above some critical density which depends on $\beta$. Larger $\beta$ leads to larger deviations from GR, but notice that the exponential theory has more of an effect than the hyperbolic theory does. \textbf{Right:} Scalar charge as a function of the total baryonic mass. The ``spontaneous'' activation of the scalar field occurs at a particular mass, while the dependance on $\beta$ appears to be independent of the theory. Again we see that the exponential theory causes a larger activation of the scalar field than the hyperbolic theory.
		}
	\end{figure*}

	We numerically solve the system of equations above parameterized by central density $\rho_c$ and coupling constant $\beta$. Following the methods employed in Ref.~\cite{Anderson:2016fi} we use \texttt{Mathematica}'s default integrator which makes use of an LSODA approach (a variant of the Livermore Solver for Ordinary Differential Equations)~\cite{mathematica}. Inside the star the pressure is governed by the piecewise polytropic EoS, while the pressure vanishes outside the star with the boundary where the pressure vanishes marking the surface of the star. We integrate the equations from the center of the star to an effective infinity that is far from the surface of the star, which allows us to determine the metric and the scalar field in the entire spacetime. Far from the star, the scalar field behaves as
	\be
		\varphi \eq \varphi_\infty + \dfrac{\omega}{r_*} + {\cal{O}}\left(\frac{1}{r_{*}^{2}}\right)\,\,,
	\label{eq:scakar_inf}
	\ee
	where $\varphi_\infty$ is the asymptotic value of the scalar field determined from cosmology (which we fix to a small value and enforce as a boundary condition) and $\omega$ is a constant that is proportional to the scalar charge. We define the scalar charge via
	\be
		\alpha_{\text{\footnotesize sc}} \eq \dfrac{\omega}{Gm_*}\,\,,
	\label{eq:charge}
	\ee
	where $m_*$ is the Einstein-frame ADM mass of the star and the subscript ``sc'' is used to distinguish the scalar charge from the conformal coupling $\alpha(\varphi)$. Physically, the scalar charge is a measure of the effective coupling between the scalar field and the star, with a larger value meaning stronger coupling. The scalar charge is then determined by $\omega$ which we find by extracting the $1/r$ piece of the external solution of the scalar field.
	
\subsection{$\beta<0$: Standard Spontaneous Scalarization}

	Choosing a sufficiently negative value of $\beta$ ($\lesssim -4$ for most EoS) leads to an instability in the star that causes an activation of the scalar field above its background cosmologically-determined value, a process known as spontaneous scalarization~\cite{Damour:1992we,Damour:1993hw,Damour:1996ke}. In the context of cosmology, however, all theories with $\beta<0$ generically lead to drastic disagreement with Solar System observations~\cite{Anderson:2016fi,Sampson:2014qqa}, at least for the subset of symmetric, monotonically increasing coupling potentials that we consider here. Thus, the NS solutions presented in this section should be strictly seen as a stepping stone towards giving a better understanding of what will be presented for the $\beta>0$ cases.
	
	Figure~\ref{fig:negative_beta} shows the results of our numerical calculations using both exponential and hyperbolic couplings for $\beta=-6$ and $\beta=-5.5$ as examples. For these calculations we employ the ENG piecewise polytropic EoS with parameters given in Table~\ref{tab:eos}, but it should be noted that the qualitative behavior seen here is general among all EoS that we consider. At low densities, only the GR solution exists and the scalar field remains at its background value. There exists a critical density at which the scalar field begins to ``turn on'' and produces sudden deviations from GR as can be seen in the left panel of Fig.~\ref{fig:negative_beta}. The critical density at which this happens depends on the EoS and the value of $\beta$, but as we stated earlier, it is a generic feature among all EoS considered here. Larger central densities eventually lead back to the GR solution and a trivial scalar field profile, just as in the low density case. 

	A larger value of $\beta$, which in these theories explicitly means a stronger coupling to matter, leads to larger deviations from GR and therefore larger values of scalar charge $\alpha_\SC$ (cf. the right panel of Fig.~\ref{fig:negative_beta}). This is consistent with the idea that scalar charge is the effective coupling between the star and the scalar field. Notice, however, that the exponential theory predicts larger deviations from GR than the hyperbolic theory does (i.e.~the charge is roughly three times larger), regardless of what value of $\beta$ we choose. This can be understood by looking at the first term in square bracket in Eq.~(\ref{eq:NSstructure_psi}), or more explicitly the fact that $\psi' \propto \alpha(\varphi)$, which provides a type of feedback loop. In the exponential theory, large values of $\varphi$ produce large values of $\alpha(\varphi)$ relative to the hyperbolic theory, which is obvious from the right panel of Fig.~\ref{fig:potentials_couplings}. Larger values of $\alpha(\varphi)$ thus lead to a larger positive feedback into $\psi'$ and so on, explaining why the exponential theory leads to larger activation of the scalar field for given values of central density.
	

	It is useful to provide an analytic explanation behind this spontaneous activation of the scalar field. In both theories, $\alpha(\varphi) = \beta \varphi + {\cal{O}}(\varphi^{2})$ for small values of $\beta\varphi$, which must be true due to the arguments presented in Sec.~\ref{Cosmological Evolution and solar system Constraints}. Following arguments and notation presented in Refs.~\cite{Damour:1993hw,Anderson:2016fi}, consider then the evolution of the scalar field in a weakly gravitating, spherically symmetric system such that $\Box \rightarrow \delta_{ij} \nabla_i \nabla_j \rightarrow \nabla^2_r$ , where $\nabla^2_r$ is the radial part of the Laplacian in spherical coordinates. For simplicity we will assume $T^*_\mat$ is constant and while weakly gravitating systems would be expected to have $T^*_\mat<0$ we allow it to remain arbitrary in the interest of generality. These assumptions reduce the field equation for the scalar field to 
	\be
		\nabla^2_r \varphi \eq -K^2\varphi \,\text{sign}(\beta T^*_\mat)\,\,,
	\label{eq:KGsimple}
	\ee
	where $K^2 = (\kappa/2)|\beta T^*_\mat|$ for $r_*<R$ and must vanish outside the star. The solution to this must still obey the physical boundary conditions of the scalar field, i.e. $\varphi(0)=\varphi_c$ and $\varphi'(0) =0$ to ensure regularity at the center and the solution must be continuous and differentiable at the surface $r_*=R$. When the product $\beta T^*_\mat$ is positive the solution becomes 
	\be
		\varphi \eq \dfrac{\varphi_\infty}{\cos(KR)}\dfrac{\sin(Kr)}{Kr}\,\,.
	\label{eq:simple_solution}
	\ee

	Even when $\varphi_\infty \approx 0$, which we emphasize is the case here, a nontrivial scalar-field solution can arise when $KR \approx \pi/2$. This type of ``resonance'' between the scalar field and matter explains why spontaneous scalarization occurs in DEF-like theories and why there exist deviations from GR in Fig.~\ref{fig:negative_beta}. However, when $\beta T^*_\mat$ is negative the scalar field is exponentially suppressed to its background value, which can be seen by replacing the trig functions in Eq.~(\ref{eq:simple_solution}) with their corresponding hyperbolic counterparts. It is then reasonable to expect that only certain ranges of density, which ultimately controls $T^*_\mat$, will lead to an activation of the scalar field for any given $\beta$, and why for relatively small and large central densities we only see the GR solution in Fig.~\ref{fig:negative_beta}.
	
	We here focused explicitly on the the sign of $\beta T^*_\mat$ rather than just $\beta$ for a reason. Indeed, for the weakly gravitating system used in the example above $T^*_\mat<0$ and the sign in Eq.~(\ref{eq:simple_solution}) would be completely determined by the sign of $\beta$. However, cosmological constraints force $\beta>0$, so unless $T^*_\mat$ were also positive one would not expect scalarization to occur. This was studied recently in Refs~\cite{Mendes:2016fby,Mendes:2015gx,2016PhRvD..93d4009P} and will be the focus of the next section.

\subsection{$\beta >0$: Go Big or Go Home}

	Studying NSs in STTs with $\beta>0$ is important because this is the region of parameter space that is consistent with Solar System constraints upon cosmological evolution of the scalar field. Luckily, NSs are extremely dense and there can exist regions near the core where $T^*_\mat$ is indeed positive, allowing for possible scalarized branches of solutions. Reference~\cite{Mendes:2016fby}, however, demonstrated that NSs that scalarize in the exponential theory lead to gravitational collapse. We will start our discussion by summarizing the results presented in Refs.~\cite{Mendes:2016fby,2016PhRvD..93d4009P} and then present the results of our numerical study of NSs in the hyperbolic theory with $\beta>0$.
	
\subsubsection{Gravitational Collapse in Exponential Theory}

\begin{figure}[h]
		\centering
		\includegraphics[width=3.5in]{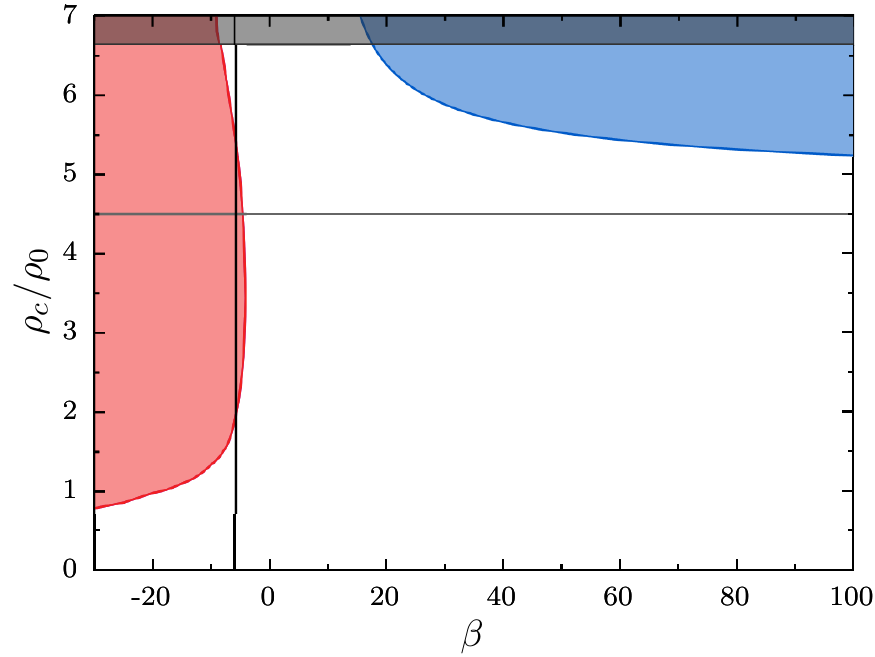}
		\caption{ \label{fig:instability} (Color Online) A rough schematic representation of the instability regions from linear stability analysis carried out in Ref.~\cite{Mendes:2016fby}. Red/Blue regions are unstable to scalar-field perturbations and the black region corresponds to solutions that are unstable to gravitational collapse. The horizontal line marks the density above which $T^*_\mat >0$ at the center of the NS. We mark a vertical line at $\beta=-6$ (corresponding to an example scenario we used in Fig.~\ref{fig:negative_beta}) to help demonstrate the behavior of the instability.
		}
	\end{figure}

	The instability of the exponential theory can be understood through a linear stability analysis of Eq.~(\ref{eq:KG}). By allowing $\varphi = \varphi_0 + \delta \varphi$ and keeping terms up to first order one finds
	\be
		\square_*\,\delta\varphi \eq -\dfrac{\kappa}{2}\beta_0T^*_\mat\,\delta\varphi\,\,,
	\label{eq:linear_pert}
	\ee
	where $\beta_0 = (\partial \alpha/\partial\varphi)_{\varphi_0}$ is like the curvature of the coupling potential\footnote{Notice that Eq.~(\ref{eq:linear_pert} encompasses a large class of STTs, not just the exponential model. This analysis applies to STTs whose coupling functions $\alpha(\varphi)$ can be approximated by $\beta \varphi + \mathcal{O}(\varphi^2)$.}, and the box operator and the stress-energy tensor are those of GR. Reference~\cite{Mendes:2016fby} performed a numerical search for solutions to Eq.~(\ref{eq:linear_pert}) of the form $\delta\varphi = e^{\Omega t} f(r)$ with a spherically symmetric background, a perfect fluid stress-energy tensor, and a polytropic EoS (in general, the location of the instability regions will depend on the EoS used). The results of this search are presented in Fig.~2 of Ref.~\cite{Mendes:2016fby} and we provide a \emph{rough} schematic drawing of those results in Fig.~\ref{fig:instability}.\footnote{Figure~\ref{fig:instability} is indeed a quantitatively replication of the results formally presented in Ref.~\cite{Mendes:2016fby}. We adjusted the regions to match those of the ENG EoS studied here. We only present this figure here for completeness and to act as a visual aid to explain the behavior of the scalar instability.} The result of the stability analysis are independent of the theory being studied and only depend on the value of $\beta_0$ defined above.

	The red/blue regions in Fig.~\ref{fig:instability} correspond to regions that are unstable to scalar field perturbations, which could lead to a non-trivial scalar-field profile inside the star. White regions, on the other hand, are regions in the parameter space where only the GR solution is present and no excitations of the scalar field would be expected inside the star. We mark a line at $\beta=-6$ to clearly show that as central density increases the instability arises but then as density continues to rise the instability eventually turns off. The turn on/off of the instability in the $\beta<0$ regime is consistent with the results we previously presented in Fig.~\ref{fig:negative_beta} for both theories. Recall that the instability is related to the sign of $(\beta T^*_\mat)$ as we argued in the previous section. For small $\rho_c$ and $\beta<0$ the sign of $(\beta T^*_\mat)$ is positive, causing the amplification of the scalar field inside the star. As $\rho_c$ grows larger eventually $T^*_\mat$ becomes positive, causing $(\beta T^*_\mat)$ to become negative, ultimately suppressing the scalar field rather than amplifying it, which is why the instability eventually vanishes as $\rho_c$ increases. 
	
	In the $\beta>0$ region of the parameter space the behavior of the instability is slightly different. The instability to scalar perturbations no longer vanishes for larger values of $\rho_c$, meaning that once the central density becomes large enough, i.e. in the blue regions of Fig.~\ref{fig:instability}, the instability will be present inside the star. The work in Refs.~\cite{Mendes:2016fby,2016PhRvD..93d4009P} showed that the final fate of scalarized NSs existing in the $\beta>0$ region of  Fig.~\ref{fig:instability} in the exponential theory was always gravitational collapse. Our understanding of these results is that NSs can exist in the exponential theory with $\beta>0$, provided that they live outside of the blue region of Fig.~\ref{fig:instability} and are identical to the trivial GR solutions. However, since these NSs would be identical to NSs in GR in these regions of parameter space no scalarized solutions would exist in nature. Because NSs in the exponential theory that could undergo scalarization would ultimately collapse to black holes we focus our attention to the study of NSs in the hyperbolic theory.  
	
\subsubsection{Scalarization in the Hyperbolic Theory}
\begin{figure}[h]
		\centering
		\includegraphics[width=3.5in]{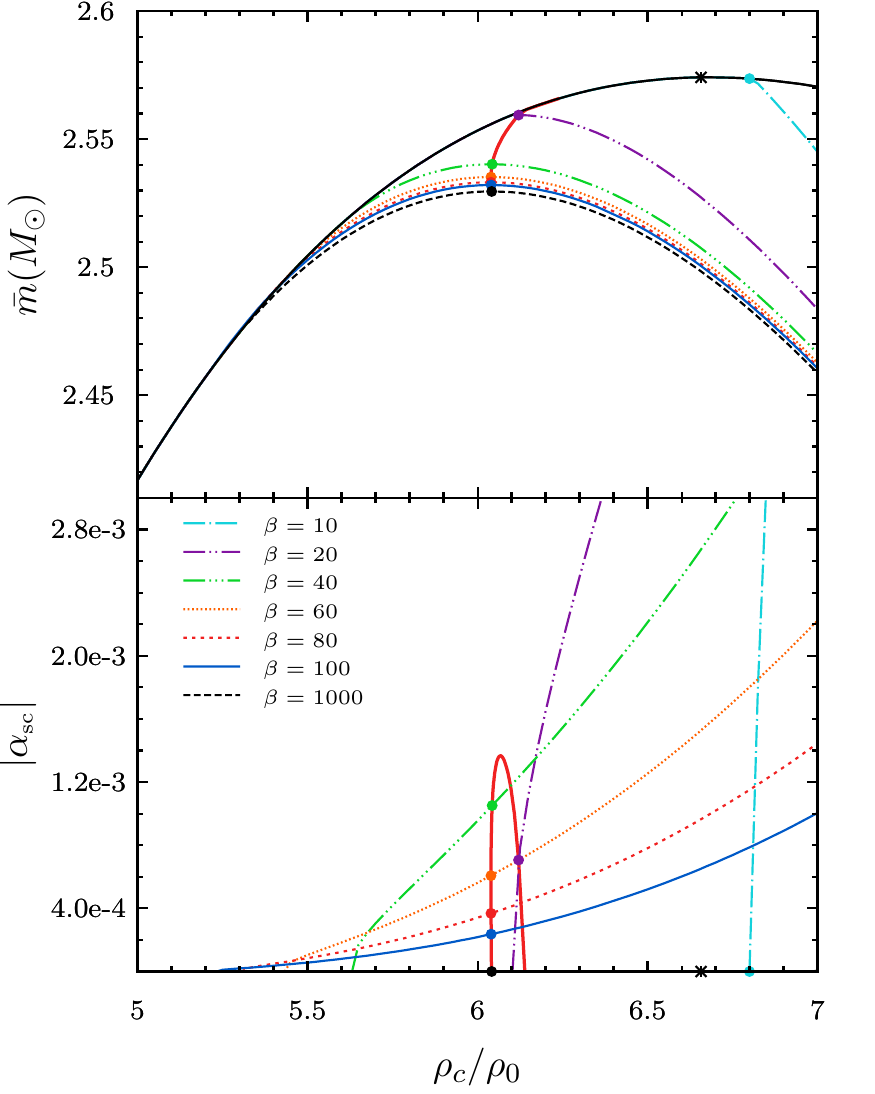}
		\caption{ \label{fig:COSH_Mrho} (Color Online) \textbf{Top:} Baryonic mass as a function of central baryonic density for a wide range of $\beta$. There exists an asymptotic solution as $\beta \rightarrow \infty$ (as $\beta$ increases the branches only experience small changes from one to the next). Solid points indicate the maximum mass that is stable and the solid red line  connects all maximum mass solutions for all $\beta$. \textbf{Bottom:} The scalar charge as a function of the central baryonic density for the same set of $\beta$'s as in the top panel. Solid points here correspond to the same solutions at the solid points in the top panel, with the solid red curve showing how these values change as a function of $\beta$. Even though the scalar charge grows monotonically there exists a maximum value that it can take for stable NS solutions. For large $\beta$, increasing $\beta$ continues to decrease the maximum stable scalar charge, though it does not cause the charge to vanish.
		}
	\end{figure}

	Following the numerical procedure explained at the beginning of the section, we calculate NS solutions in all 5 EoSs given in Table~\ref{tab:eos} for $0\leq \beta \lesssim 450$ and $\beta=1000$. While we do not analyze the dynamical stability of the static solutions we find, we define stability based on a turning point criterion in a sequence of equilibrium configuration described in detail in Ref.~\cite{Harada:342978,1997PThPh..98..359H}. In short, every solution to the left of the maximum mass in a mass-density plot, like that in Fig.~\ref{fig:COSH_Mrho}, would be considered stable and every solution to the right would be unstable.  Figure~\ref{fig:COSH_Mrho} shows our numerical solutions using the ENG EoS for a small subset of $\beta$, with the top panel demonstrating stronger deviations from GR for larger values of $\beta$. There are three key features of note in these solutions: %
	\begin{enumerate}
		\item there exist values of $\beta$ that have no stable branch of solutions,
		\item there appears to be an asymptotic branch of solution for large values of $\beta$,
		\item there will exist a maximum mass and stable scalar charge $\alpha_{\SC,\mbox{\tiny max}}$ for each $\beta$.
	\end{enumerate}

	Let us investigate the first point above. From our turning point definition of stability, any solution existing on a branch to the right of the maximum will be unstable to gravitational collapse. Because small values of $\beta$ do not lead to deviations from GR until after the maximum mass is reached, it stands to reason that there exists a $\beta_{\mbox{\tiny min}}$ such that only values of $\beta>\beta_{\mbox{\tiny min}}$ allow for stable scalarized NS solutions in the hyperbolic theory. For example, in the top panel of Fig.~\ref{fig:COSH_Mrho} the $\beta=10$ curve clearly  does not have stable solutions different from GR. In other words, sufficiently small values of $\beta$ only lead to deviations from GR when the $\rho_c$ is large and in the black region of Fig.~\ref{fig:instability} and would undergo gravitational collapse. When $\beta=40$ there exists a clear set of solutions different from GR that would indeed be stable up to the turning point, which we indicate with a circle in the figure for clarity. The values of $\beta_{\mbox{\tiny min}}$ are listed in Table~\ref{tab:eos} for all EoSs studied.
	
	There also appears to be an asymptotic branch of solutions in the limit $\beta \rightarrow \infty$, a qualitative feature occurring in each EoS studied. Starting from $\beta=0$, increasing $\beta$ initially induces dramatic changes in the corresponding branch of solutions; however, for $\beta \gtrsim 40$ for the ENG EoS, the relative change between solution branches changes only slightly even when $\beta$ is increased by an order of magnitude. This idea can be explained by again referring back to the structure of $\alpha(\varphi)$ in Fig.~\ref{fig:potentials_couplings}. As $\beta \rightarrow \infty$ we find that $\alpha(\varphi) \rightarrow 1/\sqrt{3}$ exponentially. Therefore, as long as $\varphi$ does not identically vanish everywhere, the coupling to matter is saturated, and thus, there will exist a branch of solution representing this maximal coupling case, which we see in Fig.~\ref{fig:COSH_Mrho} as $\beta$ increases. This is a feature absent in the exponential theory because $\alpha(\varphi)$ is linear in $\beta$ and therefore it has no asymptotic limit, meaning that the scalar field's coupling to matter can never be saturated.
	
	Stability arguments~\cite{Harada:342978,1997PThPh..98..359H} indicate that there exists a maximum mass, and therefore a maximum stable scalar charge, on each branch of solutions determined by $\beta$. However, the bottom panel of Fig.~\ref{fig:COSH_Mrho} indicates that the scalar charge grows monotonically as the central density increases. Using the turning point criterion allows us to extract the most massive stable solution and hence the corresponding scalar charge of that solution. The points marked in Fig.~\ref{fig:COSH_Mrho} correspond to identical solutions in each panel and for $\beta\gtrsim 40$ we see that the maximum stable scalar charge on each solution branch decreases with increasing $\beta$. There actually exists a $\beta_{\mbox{\tiny max}}$ where $\alpha_{\SC,\mbox{\tiny max}}(\beta)$ peaks, which is evident in Fig~\ref{fig:max_charge_beta} and whose values appear in Table~\ref{tab:eos}.

\begin{figure}[h]
		\centering
		\includegraphics[width=3.5in]{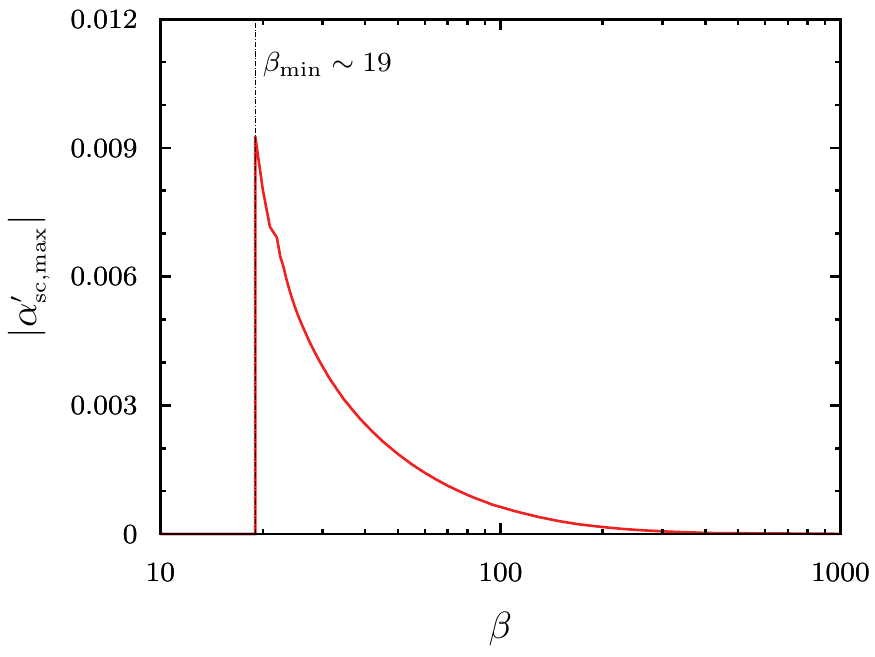}
		\caption{ \label{fig:COSH_alpha'} (Color Online) Rate of change of the scalar charge evaluated at the maximum mass solutions (solutions identified by points in Fig.~\ref{fig:COSH_Mrho}) as a function of $\beta$ for the ENG EoS. There are no values for $\beta\lesssim 19$ because there are no stable solutions with scalar charge, c.f. the discussion in the text. Notice that the ``rapidity'' of the increase of scalar charge monotonically decreases with increasing $\beta$.
		}
	\end{figure}

	Figure~\ref{fig:max_charge_beta} shows the maximum stable scalar charge as a function of $\beta$ for all EoSs in Table~\ref{tab:eos}. There is a general trend that EoSs that allow for larger compactness have large values of scalar charge. This complements the results presented in Ref.~\cite{Mendes:2015gx} where the author demonstrated a relation between the compactness and the onset of scalarization for the same set of EoSs. We can also roughly quantify how rapid the scalar field ``turns on'' \footnote{This activation, or ``turn on'', is not dynamical. We are simply referring to how rapid the charge increases as $\rho_c$ increases.} by measuring the slope of the charge in the bottom panel of Fig.~\ref{fig:COSH_Mrho} near the maximum stable solutions, i.e. the ones marked with points. Figure~\ref{fig:COSH_alpha'} shows the value of this slope, which we denote $\alpha_{\SC,\mbox{\tiny max}}' = \partial \alpha_{\SC,\mbox{\tiny max}}/\partial \rho_c$, as a function of $\beta$. For small $\beta$ there is no change in the charge because, as we previously stated, there are no stable scalarized solutions below $\beta_{\mbox{\tiny min}}$. As one may expect from Fig.~\ref{fig:COSH_Mrho} there is a clear monotonic decrease in $\alpha_{\SC,\mbox{\tiny max}}'$ as we increase $\beta$. A decreasing $\alpha_{\SC,\mbox{\tiny max}}'$ means that the scalar field grows much more gradually with central pressure as $\beta$ becomes larger.
	
	To understand why the maximum scalar charge decreases for large $\beta$ we must return to the field equations, particularly Eq.~(\ref{eq:NSstructure_psi}) which we rewrite for convenience as
	\ba
		\psi' &=& \dfrac{4\pi G r A^4(\varphi)}{(r_*-2\mu)}\big[ \alpha(\varphi) (\epsilon - 3p) + r_*\psi(\epsilon - p)\big]\nn\\
		&\,&\,\,\,-2\psi\dfrac{(1-\mu/r_*)}{(r_*-2\mu)}	\,\,.\nn
	\ea
	This equation describes the curvature of the scalar field profile, with the first term in square brackets identified with $(-\alpha(\varphi) T^*_\mat)$. This term will explicitly be negative near the center of the NS because $T^*_\mat>0$. Increasing $\beta$ increases the value of $\alpha(\varphi)$, even though only slightly when $\beta$ is large. This term then contributes to a more negative value of  curvature, which causes the scalar field to decay faster inside the star as $\beta$ increases. Numerically, we find that the NS solutions appearing in Fig.~\ref{fig:max_charge_beta} have nearly identical central scalar field values for $\beta>100$. Because the scalar field decays faster inside the star there will exist a smaller value of $\omega$ from Eq.~(\ref{eq:scakar_inf}) upon matching boundary conditions at the surface and therefore a smaller scalar charge.

\section{Conclusion}\label{Conclusion}

	Our calculations show that theories with an exponential conformal factor (identical to the ones first studied by Damour and Esposito-Far\`ese~\cite{Damour:1992we,Damour:1993hw}) and theories with a hyperbolic conformal factor (motivated from quantum field theory~\cite{Mendes:2015gx,birrell1984quantum,Lima:2010dh}) always pass Solar System constraints provided that $\beta \lesssim \beta_\crit=24$ and $\beta \lesssim \beta_\crit=17$ respectively for all initial conditions consistent with BBN constraints. By considering a random distribution of initial conditions below the BBN constraint, these bounds can be relaxed to $\beta_\crit \sim 34$ and $\beta_\crit \sim 25$ respectively. These results tell us that Solar System constraints place a rather tight bound on the $\beta$ parameter if one wishes to avoid fine-tuning the initial conditions at the time of BBN.
	
	Recent work has suggested that scalarization is still possible in STTs with $\beta>0$ if the star becomes dense enough such that the trace of the matter stress-energy tensor becomes positive inside the star, which is only possible for certain EoSs~\cite{Mendes:2015gx,Mendes:2016fby,2016PhRvD..93d4009P}. However, neutron stars in STTs with an exponential conformal factor have recently been shown to undergo gravitational collapse when $\beta>0$~\cite{Mendes:2016fby}; therefore, only NSs with $T^\mat=-\rho + 3p <0$ would exist in this theory and these would not scalarize. After verifying the results in these recent studies, we constructed NSs in a theory with hyperbolic coupling and $\beta>0$ for $\beta$ regions consistent with Solar System observations. Our results show that there exists a minimum value of $\beta$, denoted $\beta_{\min}$ in Table~\ref{tab:eos}, for which there exist \emph{stable} scalarized NS solutions that are distinct from GR. However, in every EoS that we considered, all of these values of $\beta_{\min}$ were larger than the upper bound on $\beta$ placed with Solar System observations when one saturates BBN initial conditions. 
	
	The fact that $\beta_\crit<\beta_{\min}$ for many EoSs does not mean that no scalarized NSs can exist. Indeed, one can choose more finely-tuned initial conditions at the time of BBN, thus making $\beta_{\crit}$ slightly larger. The larger one wishes to make $\beta_{\crit}$, the more one will need to fine-tune the initial conditions. Eventually, such a large fine-tuning of initial conditions would have to be justified through a theoretical physics mechanism that currently does not exist. Alternatively, one can pick a value of $\beta > \beta_{\crit}$ such that Solar System constraints are satisfied today but fail in the future. In this case, the likelihood of the existence of stable, scalarized NSs becomes smaller the larger $\beta$ is above $\beta_\crit$, as the bands of allowed $\beta$ thin out and become more sparse as $\beta$ increases. 
	

	The results presented here and in Ref.~\cite{Anderson:2016fi} may beg the question of whether or not it is possible to engineer a conformal factor in massless STTs that generally passes Solar System tests and, furthermore, leads to scalarization in neutron stars. Solar System tests demand that, at least locally, the conformal coupling potential have positive curvature, i.e.~$\beta(\varphi)>0$ in general, and therefore one must have an EoS that allows $T^\mat>0$ in order to have scalarized neutron stars. It seems that having $\beta(\varphi)>0$ and $T^\mat>0$ only allows for a scalar of $\mathcal{O}(10^{-3})$, where as in the $\beta(\varphi)<0$ case the charge is of $\mathcal{O}(10^{-1})$. Therefore, these consistency between Solar System observations and the cosmological evolution of the scalar field may limit the maximum size of the scalar charge that can occur in NSs for a large class of massless STTs. A promising avenue for future research would be the study of massive STTs~\cite{2017arXiv170306341A,Pretorius:2016wp} in this context or to perform an analysis like the one in this paper on tensor-multiscalar theories. 
	
	If consistency between the bounds from Solar System tests and scalarized NSs were achieved, then NSs in these theories would have a different mass and radius than that predicted in GR and only deviate from GR after a critical density (or compactness) is achieved such that the trace of the stress-energy tensor is positive. A direct observation of such a system, perhaps with NASA's NICER telescope~\cite{2012SPIE.8443E..13G,2014SPIE.9144E..20A}, would allow further constraints to be placed on STTs. Such an analysis would require null ray tracing of light emitted at the surface of such STT deformed NSs. These types of results would provide useful information on what the actual form of the conformal factor may be. We have seen that some models, like the exponential one, do not allow for scalarized NSs, while others, like the hyperbolic one, do. Therefore, any observation of neutron stars that could shed some light on the functional form of the conformal factor would be valuable. 
		
	Similarly, it would also be interesting to investigate dynamical and induced scalarization in hyperbolic STTs with $\beta >0$ with an eye out for gravitational wave tests with ground-based instruments. Given the results obtained in the spontaneous scalarization case, we expect the scalar charge to be dynamically generated in binaries, but its magnitude to be significantly smaller than in the exponential $\beta < 0$ case. The latter case, however, is already ruled out by Solar System observations, while we showed here that hyperbolic STTs with $\beta>0$ can still be consistent. Studying scalarization in these theories would allow us to predict the signatures that would imprint on the gravitational waves emitted in neutron star binaries. Presumably, the activation of the scalar field will induce dipole radiation, which in turn will speed up the inspiral of the binary, leaving a signature on the gravitational waves emitted. If so, gravitational wave observations with advanced LIGO could place the first meaningful constraints on such theories in the context of NSs.

\acknowledgements
We thank Emanuele Berti, Enrico Barausse, Norbert Wex, Hector Okada da Silva, Raissa Mendes, and Nestor Ortiz for useful discussions on neutron stars in scalar-tensor theories, and for reading our manuscript carefully and providing useful suggestions. N.Y.~acknowledges support from the NSF CAREER grant PHY-1250636 and NASA grant NNX16AB98G. 

\bibliography{main,master}
\end{document}